\documentclass[useAMS,usenatbib]{mn2e}
\usepackage{url,times,graphicx,amsmath,amsfonts,amssymb,aas_macros,color,epsfig}

\newcommand{\Tab}[1]{Table~\ref{#1}}
\newcommand{\Sec}[1]{Section~\ref{#1}}

\newcommand{\Fig}[1]{Fig.~\ref{#1}}
\newcommand{\beq}{\begin{equation}}
\newcommand{\eeq}{\end{equation}}
\newcommand{\hMpc}{{\ifmmode{h^{-1}{\rm Mpc}}\else{$h^{-1}$Mpc}\fi}}
\newcommand{\hkpc}{{\ifmmode{h^{-1}{\rm kpc}}\else{$h^{-1}$kpc}\fi}}
\newcommand{\hMsun}{{\ifmmode{h^{-1}{\rm {M_{\odot}}}}\else{$h^{-1}{\rm{M_{\odot}}}$}\fi}}
\def\beqa{\begin{eqnarray}}
\def\eeqa{\end{eqnarray}}
\def\hMpc{$h^{-1}\,{\rm Mpc}$}
\def\hkpc{$h^{-1}\,{\rm kpc}$}

\def\vmax{$v_{\rm max}$}

\def\head{
 \vbox to 0pt{\vss
                   \hbox to 0pt{\hskip 440pt\rm LA-UR-10-07069\hss}
                  \vskip 25pt}}

\title[Halo-Finder Comparison Project]
      {Haloes gone MAD\footnote{airport code for Madrid, Spain}: The Halo-Finder Comparison Project}
\author[Knebe et al.]
       {Alexander Knebe$^{1}$\thanks{E-mail: alexander.knebe@uam.es}, Steffen R. Knollmann$^1$, Stuart I. Muldrew$^{2}$, Frazer R. Pearce$^{2}$, \newauthor
         Miguel Angel Aragon-Calvo$^{21}$,  Yago Ascasibar$^{1}$, Peter S. Behroozi$^{28,29,30}$, \newauthor
         Daniel Ceverino$^{4}$, Stephane Colombi$^{24}$,  Juerg Diemand$^{22}$, Klaus Dolag$^{11}$,  \newauthor 
         Bridget L. Falck$^{21}$,  Patricia Fasel$^{9}$, Jeff Gardner$^{18}$, Stefan Gottl\"ober$^{14}$, \newauthor 
         Chung-Hsing Hsu$^{10}$, Francesca Iannuzzi$^{11}$, Anatoly Klypin$^{5}$, Zarija Luki\'c$^{8}$,  \newauthor 
         Michal Maciejewski$^{11}$, Cameron McBride$^{19}$, Mark C. Neyrinck$^{21}$, Susana Planelles$^{3}$, \newauthor 
         Doug Potter$^{22}$,  Vicent Quilis$^{3}$, Yann Rasera$^{17}$, Justin I. Read$^{26,27}$, Paul M. Ricker$^{6,7}$,\newauthor
         Fabrice Roy$^{17}$, Volker Springel$^{12,13}$,  Joachim Stadel$^{22}$,  Greg Stinson$^{20}$, \newauthor
         P. M. Sutter$^{6}$,  Victor Turchaninov$^{16}$, Dylan Tweed$^{23}$, Gustavo Yepes$^{1}$, \newauthor 
         Marcel Zemp$^{25}$\\
\\
$^{1}$Departamento de F\'isica Te\'orica, M\'odulo C-15, Facultad de Ciencias, Universidad Aut\'onoma de Madrid, 28049 Cantoblanco, Madrid, Spain\\
$^{2}$School of Physics \& Astronomy, University of Nottingham, Nottingham, NG7 2RD, UK\\
$^{3}$Departament   d'Astronomia   i   Astrof\'{\i}sica,   Universitat de Val\`encia,   46100  -   Burjassot  (Valencia),   Spain\\
$^{4}$Racah Institute of Physics, The Hebrew University, Jerusalem 91904, Israel\\
$^{5}$Department of Astronomy, New Mexico State University, Las Cruces, NM 88003-0001, USA\\
$^{6}$Department of Physics, University of Illinois at Urbana-Champaign, Urbana, IL 61801-3080, USA\\
$^{7}$National Center for Supercomputing Applications, University of Illinois at Urbana-Champaign, Urbana, IL 61801, USA\\
$^{8}$T-2, Theoretical Division, Los Alamos National Laboratory, P.O. Box 1663, Los Alamos, NM 87544, USA\\
$^{9}$CCS-3, Computer, Computational and Statistical Sciences Division, Los Alamos National Laboratory, P.O. Box 1663, Los Alamos, NM 87544, USA\\
$^{10}$Computer Science and Mathematics Division, Oak Ridge National Laboratory, P.O. Box 2008, Oak Ridge, TN 37831, USA\\
$^{11}$Max-Planck Institut f\"{u}r Astrophysik, Karl-Schwarzschild Str. 1, D-85741 Garching, Germany\\
$^{12}$Heidelberg Institute for Theoretical Studies, Schloss-Wolfsbrunnenweg 35, 69118 Heidelberg, Germany\\ 
$^{13}$Zentrum f\"ur Astronomie der Universit\"at Heidelberg, ARI, M\"{o}nchhofstr. 12-14, 69120 Heidelberg, Germany \\
$^{14}$Astrophysikalisches Institut Potsdam, An der Sternwarte 16, 14482 Potsdam, Germany\\
$^{16}$Keldysh Institute of Applied Mathematics, Russian Academy of Sciences, 125047 Moscow, Russia\\
$^{17}$CNRS, Laboratoire Univers et Th\'eories (LUTh), UMR 8102 CNRS, Observatoire de Paris, Universit\'e Paris Diderot; 5 Place  Jules Janssen, 92190 Meudon, France\\
$^{18}$University of Washington, Department of Physics,  Box  351560, Seattle, WA, USA\\
$^{19}$Vanderbilt University, Department of Physics \&  Astronomy, 6301 Stevenson Center, Nashville, TN 37235, USA\\
$^{20}$Jeremiah Horrocks Institute, University of Central Lancashire, Preston PR1 2HE, UK\\
$^{21}$Department of Physics and Astronomy, Johns Hopkins University, 3701 San Martin Drive, Baltimore, MD 21218, USA\\
$^{22}$University of Zurich, Institute for Theoretical Physics, Winterthurerstrasse 190, CH-8057 Zurich, Switzerland\\
$^{23}$Institut d'Astrophysique Spatiale, CNRS/Universite Paris-Sud 11, 91405 Orsay, France\\
$^{24}$Institut d'Astrophysique de Paris, CNRS UMR 7095 and UPMC, 98bis, bd Arago, F-75014 Paris, France\\
$^{25}$University of Michigan, Department of Astronomy, 500 Church
St., Ann Arbor, MI, 48109-1042, USA\\
$^{26}$Institute for Astronomy, Department of Physics, ETH Z\"urich, Wolfgang-Pauli-Strasse 16, CH-8093 Z\"urich, Switzerland\\
$^{27}$Department of Physics and Astronomy, University of Leicester, University Road, Leicester LE1 7RH, UK\\
$^{28}$Kavli Institute for Particle Astrophysics and Cosmology, Stanford, CA 94309, USA\\
$^{29}$Physics Department, Stanford University, Stanford, CA 94305, USA\\
$^{30}$SLAC National Accelerator Laboratory, Menlo Park, CA 94025, USA\\
}

\begin{document}

\date{Accepted XXXX . Received XXXX; in original form XXXX}

\pagerange{\pageref{firstpage}--\pageref{lastpage}} \pubyear{2010}

\maketitle

\label{firstpage}

\clearpage

\begin{abstract}

  We present a detailed comparison of fundamental dark
  matter halo properties retrieved by a substantial number of
  different halo finders.  These codes span a wide range of techniques
  including friends-of-friends (FOF), spherical-overdensity (SO) and
  phase-space based algorithms. We further introduce a robust (and
  publicly available) suite of test scenarios that allows halo
  finder developers to compare the performance of their codes against
  those presented here. This set includes mock haloes containing
  various levels and distributions of substructure at a range of
  resolutions as well as a cosmological simulation of the large-scale
  structure of the universe.

  All the halo finding codes tested could successfully recover the
  spatial location of our mock haloes. They further returned lists of
  particles (potentially) belonging to the object that led to
  coinciding values for the maximum of the circular velocity profile
  and the radius where it is reached. All the finders based in
  configuration space struggled to recover substructure that was
  located close to the centre of the host halo and the radial
  dependence of the mass recovered varies from finder to finder. Those
  finders based in phase space could resolve central substructure
  although they found difficulties in accurately recovering its
  properties. Via a resolution study we found that most of the finders
  could not reliably recover substructure containing fewer than 30-40
  particles. However, also here the phase space finders excelled by
  resolving substructure down to 10-20 particles. By comparing the
  halo finders using a high resolution cosmological volume we found
  that they agree remarkably well on fundamental properties of
  astrophysical significance (e.g. mass, position, velocity, and peak
  of the rotation curve).

  We further suggest to utilize the peak of the rotation curve \vmax\
  as a proxy for mass given the arbitrariness in defining a proper
  halo edge.

\end{abstract}

\begin{keywords}
  methods: $N$-body simulations -- galaxies: haloes -- galaxies:
  evolution -- cosmology: theory -- dark matter
\end{keywords}

\section{Introduction} \label{sec:introduction}

While recent decades have seen great progress in the understanding and
modelling of the large- and small-scale structure of the Universe by
means of numerical simulation there remains one very fundamental
question that is yet to be answered: ``How to find a dark matter
halo?''  The comparison of any cosmological simulation to
observational data relies upon reproducibly identifying ``objects''
within the model.  But how do we identify ``dark matter haloes'' or
even ``galaxies'' in such simulations? Researchers in the field have
developed a wide variety of techniques and codes to accomplish this
task.  But how does the performance of these various techniques and
codes compare? While we still may argue about the proper definition of
an ``object'' the various approaches should nevertheless agree once
the same recipe for defining a (dark matter) halo is used.

This introduction begins by establishing why it is important to have
``The Halo-Finder Comparison Project'' before continuing by laying out
the groundwork for the comparison we have undertaken. It is therefore
subdivided into a first subsection where we highlight the necessity
for such a comparison and summarise the recent literature in this
area. This section also includes a brief primer on halo finders and
their history. The second part introduces the design of the test
cases, illustrated with some analysis. The last part then raises the
question ``how to cross-compare haloes?'' as well as ``what is
actually a halo?'' and presents a possible answer the authors agreed
upon.

\subsection{The Necessity for a Comparison Project}

Over the last 30 years great progress has been made in the development
of simulation codes that model the distribution of dissipationless
dark matter while simultaneously following the (substantially more
complex) physics of the baryonic component that accounts for the
observable Universe. Nowadays we have a great variety of highly
reliable, cost effective (and sometimes publicly available) codes
designed for the simulation of cosmic structure formation
\citep[e.g.][]{Couchman95, Pen95, Gnedin95, Kravtsov97, Fryxell00,
  Bode00, Springel01, Knebe01, Teyssier02, OShea04, Quilis04,
  Dubinski04, Merz05, Springel05, Bagla09, Springel10, Doumler10}.

\begin{figure}
  \psfig{figure=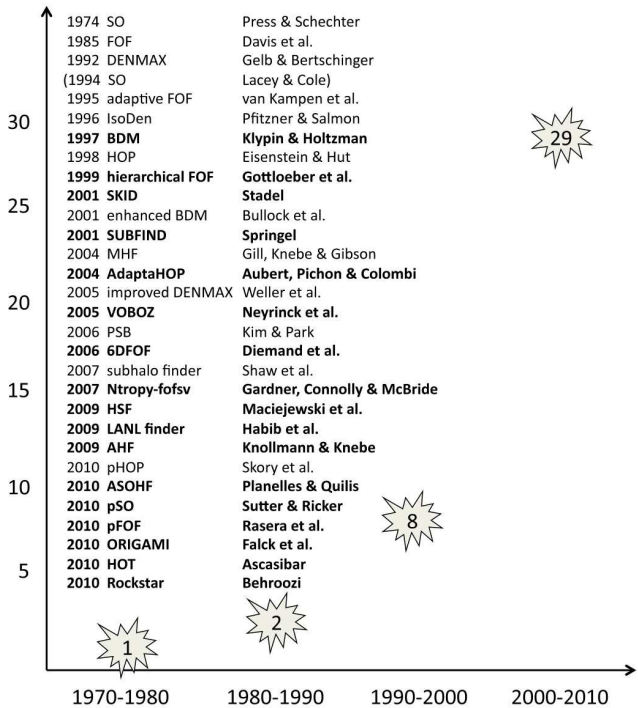,width=\hsize,angle=0}
  \caption{Schematic presentation of the (cumulative) number of halo
    finders as a function of time, binned in ten-year intervals since
    1970. The codes participating in this comparison project have been
    highlighted in bold font.}
\label{fig:HaloFinder}
\end{figure}

However, producing the (raw) simulation data is only the first step in
the process; the model requires reduction before it can be compared to
the observed Universe we inhabit. This necessitates access to analysis
tools to map the data onto ``real'' objects; traditionally this has
been accomplished via the use of ``halo finders''. Conventional halo
finders search the (dark) matter density field within the simulations
generated by the aforementioned codes to find locally over-dense
gravitationally bound systems, which are then tagged as (dark) matter
haloes. Such tools have led to critical insights into our
understanding of the origin and evolution of cosmic structure. To take
advantage of sophisticated simulation codes and to optimise their
predictive power one obviously needs equally sophisticated halo
finders!  Therefore, this field has also seen great development in
recent years \citep[e.g.][see also \Fig{fig:HaloFinder}, noting that
for some halo finders no code paper exists yet]{Gelb94, Klypin97,
  Eisenstein98hop, Stadel01, Bullock01, Springel01subfind, Aubert04,
  Gill04a, Weller05, Neyrinck05, Kim06, Diemand06, Shaw07, Gardner07a,
  Gardner07b, Maciejewski09, Habib09, Knollmann09, Ascasibar10,
  Behroozi10, Planelles10, Sutter10, Rasera10, Skory10, Falck11}. But
so far comparison projects have tended to focus on the simulation
codes themselves rather than the analysis tools.

The increasing demand and supply for halo finders is schematically presented in
\Fig{fig:HaloFinder} where we show the (cumulative) number of codes as
a function of time, binned in ten year intervals since 1970. We can
clearly see the increasing pace of development in the past decade
reflecting the necessity for sophisticated codes: in the last ten
years the number of existing halo finding codes has practically
tripled. While for a long time the spherical overdensity method first
mentioned by \citet[SO, ][]{Press74} as well as the friend-of-friends
algorithm introduced by \citet[FOF, ][]{Davis85} remained the standard
techniques, the situation changed in the 90's when new methods were
developed \citep{Gelb92, Lacey94, vanKampen95, Pfitzner96, Klypin97,
  Eisenstein98hop, Gottloeber99}.

While the first generation of halo finders primarily focused on
identifying isolated field haloes the situation dramatically changed
once it became clear that there was no such thing as ``overmerging'',
i.e. the premature destruction of haloes orbiting inside larger host
haloes \citep{Klypin99} was a numerical artifact rather than a real
physical process. Now codes faced the challenge of finding both haloes
embedded within the (more or less uniform) background density of the
Universe as well as subhaloes orbiting within a density gradient of a
larger host halo. The past decade has seen a substantial number of
codes and techniques introduced in an attempt to cope with this
problem \citep{Stadel01, Bullock01, Springel01subfind, Aubert04,
  Gill04a, Weller05, Neyrinck05, Kim06, Diemand06, Shaw07, Gardner07a,
  Gardner07b, Maciejewski09, Knollmann09, Planelles10}. Along with the
need to identify subhaloes simulations became much larger during this
period and this led to a drive towards parallel analysis tools. The
simulation data had become too large to be analysed on single CPU
architectures and hence halo finders had to be developed to cope with
this situation, too.

Nevertheless, the first two halo finders mentioned in the literature,
i.e. the spherical overdensity (SO) method \citep{Press74} and the
friends-of-friends (FOF) algorithm \citep{Davis85} remain the
foundation of nearly every code: they often involve at least one phase
where either particles are linked together or (spherical) shells are
grown to collect particles. While we do not wish to invent stereotypes
or a classification scheme for halo finders there are unarguably two
distinct groups of codes:

\begin{itemize}
 \item density peak locator (+ particle collection)
 \item particle collector
\end{itemize}

\noindent
The density peak locators -- such as the classical SO method -- aim at
identifying by whatever means peaks in the matter density field. About
these centres (spherical) shells are grown out to the point where the
density profile drops below a certain pre-defined value normally
derived from a spherical top-hat collapse. Most of the methods
utilising this approach merely differ in the way they locate density
peaks. The particle collector codes -- above all the FOF method --
connect and link particles together that are close to each other
(either in a 3D configuration or in 6D phase-space). They afterwards
determine the centre of this mass aggregation.

After the initial selection has been made most methods apply a
pruning phase where gravitationally unbound particles are removed from
the object. While this unbinding procedure is not essential for
isolated field haloes it is vital for subhaloes in order to properly
alleviate contamination by host halo particles. Furthermore, for
subhaloes it appears essential to define the first guess for bound
particles upon a stable and reproducible criterion for the subhalo
edge. One cannot extend the (spherical) shells out to the point where
the density drops below some preselected multiple of the universal
background density as this level will not be reached anymore; one
needs to ``truncate'' the object beforehand, usually at the point
where the density rises again due to the fact that the subhalo is
embedded within a host. Similarly, particle collecting codes which use
simple ``proximity'' as a criterion for grouping particles need to
adjust their yardsticks.  However, the situation may be a bit more
straightforward for 6D phase-space finders as we expect
the velocity distributions of the host and the subhalo to be
different.

Driven by the explosion of high-quality observational data,
simulations of cosmological structure formation have moved to
increasingly high mass and force resolution. The simulation codes and
techniques have been continuously refined over the past few decades
providing us with methods that are akin yet different: they all have
to solve the collisionless Boltzmann equation simultaneously with
Poisson's equation and the equations that govern gas physics. In order
to verify their credibility the past few years have seen substantial
efforts to inter-compare the results stemming from these different
techniques \citep[cf.][]{Frenk99, Knebe00, OShea05, Agertz07,
  Heitmann08cccp, Tasker08}.  \textit{However, to date the literature
  lacks a quantitative comparison of the various halo finding
  techniques.} While some efforts have been directed towards this goal
\citep[e.g.][]{Lacey94, White02, Gill04a, Cohn08, Lukic09, Tweed09, Maciejewski09,
  Knollmann09} these studies primarily scratched the surface and
no-one has yet presented a conclusive inter-comparison based upon a
well defined test suite. In addition, we would like to stress again
that the analysis of massive state-of-the-art simulations is a
non-trivial task, especially when it comes to the detailed
substructure of the haloes. Furthermore, various definitions of the
extent of a halo exist within the literature making comparisons of the
results from different groups far from straightforward
\citep[cf.][]{White01, Lukic09}.

We though acknowledge that there is a body of literature available
that has compared halo finder methods to theoretical predictions
\citep[e.g.][]{Press74,Lacey94,Sheth99,Jenkins01,Robertson09,Courtin10}. While
this is important work, it nevertheless rather often leads to halo
finders being tuned to match theoretical expectations than testing the
validity of the code in the first place; the theories have sometimes
been used to answer ``what halo definition is required to match
theoretical expectations?''. This may therefore mask important
differences between simple linear theory and the full non-linear
growth of structure in the Universe. In this paper, we focus instead
on directly comparing different codes for halo finding and leave
theoretical expectation aside.

In summary, there is no clear definition of ``what is a (dark) matter
halo?'' never mind ``what is a subhalo?''. Workers in the field of
simulation analysis tend to utilise their own definitions and codes to
study the properties of haloes in cosmological simulations. This paper
aims at rectifying this situation by presenting the first ever
coherent halo-finder comparison involving a substantial number of
codes as well as providing the community with a well-defined set of
test cases. However, we would like to caution the reader that the prime
objective of this comparison is \textit{codes} and not
\textit{algorithms}. Therefore, while certain codes may be based upon
the same algorithm they still may yield (marginally) different results
due to the individual realisation of that algorithm.

\subsection{The Workshop}
During the last week of May 2010 we held the workshop ``Haloes going
MAD'' in Miraflores de la Sierra close to Madrid dedicated to the
issues surrounding identifying haloes in cosmological
simulations. Amongst other participants 15 halo finder representatives
were present. The aim of this workshop was to define (and use!) a
unique set of test scenarios for verifying the credibility and
reliability of such programs. We applied each and every halo finder to
our newly established suite of test cases and cross-compared the
results.

To date most halo finders were introduced (if at all) in their
respective code papers which presented their underlying principles and
subjected them to tests within a full cosmological environment
(primarily matching (sub-)halo mass functions to theoretical models
and fitting functions) and hence no general benchmarks such as the
ones designed at the workshop and presented below existed prior to our
meeting. Our newly devised suite of test cases is designed to be
simple yet challenging enough to assist in establishing and gauging
the credibility and functionality of all commonly employed halo
finders.  These tests include mock haloes with well defined properties
as well as a state-of-the-art cosmological simulation. They involve
the identification of individual objects, various levels of
substructure, and dynamically evolving systems. The cosmological
simulation has been provided at various resolution levels with the
best resolved containing a sufficient number of particles (1024$^3$)
that it can only presently be analysed in parallel.

All the test cases and the analysis presented here is publicly
available from \texttt{http://popia.ft.uam.es/HaloesGoingMAD} under
the tab ``The Data''.

\subsection{How to compare Haloes?} \label{sec:howtocomparehaloes}
One of the most crucial questions to address is obviously ``How to
define a halo?''. This question is intimately related to ``How do we
fairly cross-compare the results of the various halo finders?''. While
we all agreed that the proper definition of a halo should be a
``gravitationally bound object'', how the size of a halo should be
defined proved harder to agree upon. The ``virial radius'' is not a
well-defined property as its precise definition can (and does) vary
from halo finder to halo finder.\footnote{We like to add the
  cautionary remark that a lot of the properties and in particular any
  ``radius'' is based upon the assumption of spherical symmetry which
  is not valid for all halo finders presented here.} Furthermore, this
quantity is ill-defined for subhaloes that live within the environment
of a host halo. While there is some work available that allows for a
conversion between commonly applied methods to calculate the mass of
an isolated field halo \citep[see e.g.][]{White01, Lukic09}, such
variations in definition will nevertheless lead to discrepancies in a
cross-comparison and hence we decided to abandon the ambiguous
definition for the edge of a halo and rather focus on a property that
uniquely specifies the halo for the code comparison project: the peak
of the rotation curve as characterised by \vmax\ and the radial
location of this peak $R_{\rm max}$, respectively. It has been argued
\citep[e.g.][]{Ascasibar08} that these quantities do indeed provide a
physically-motivated scale for dark matter haloes, showing that, in
contrast to the inner regions, there is substantial scatter in their
physical properties, as well as significant systematic trends with
halo mass and cosmic epoch, beyond the radius $R_{\rm max}$.

However, utilizing \vmax\ raises two obvious issues: firstly, as
\vmax\ is reached quite close to the centre of the halo its
measurement is obviously sensitive to resolution. Secondly, as the
value of \vmax\ is set by the central particles it is not very
sensitive to tidal stripping. The relationship between $R_{\rm max}$
and $R_{\rm vir}$ for a range of NFW halo concentrations is given in
figure~6 of \citet{Muldrew11}. The resolution issue can be addressed
by increasing the number of particles required when studying subhalo
properties so that \vmax\ will always be resolved sufficiently and
credibly. The relevance of the stripping issue though depends upon the
questions to be asked of the simulation data: are we interested in a
(stable) measure of the (original) infall mass of the subhalo or do we
want to quantify the mass inside the tidal radius?  For the comparison
project we decided to evaluate \vmax\ in order to have a stable
quantity. We further agreed that this quantity is better related to
observational data as it is possible to observe rotation curves (and
hence \vmax) whereas the same ambiguity applies to observers: what is
the (outer) edge of a halo and/or galaxy?  Nevertheless, we also
decided to include $N_{\rm part}$ (i.e. the total number of
gravitationally bound particles as returned by the respective halo
finder) in the comparison as a halo is (or should be) a
gravitationally bound entity. The values for $N_{\rm part}$ are the
ones directly returned by the halo finder and are based upon the
internal criteria each code uses. How (and if) to perform the
unbinding procedure and what particles to consider as belonging to the
(sub-)halo were questions left for each group taking part to answer as
they saw fit. For several groups these particle lists would normally
be pruned further during an additional post-processing phase prior to
obtaining halo properties. The numbers given here therefore serve
solely as an indicator of whether or not particles are missing and/or
-- in case of subhaloes -- belong to the host. In addition, we also
used the list of particles belonging to each halo to calculate a
fiducial $M_{200}$ value (defined via $M(<r)/4\pi
r^3=200\times\rho_{\rm crit}$) considering the object in isolation,
even for subhaloes: there are physical situations -- like the
dynamical friction on infalling loose groups \cite[e.g.][]{Read08,
  Lux10} -- where the (total) mass is the physically important
quantity. Such examples of the limitation of the \vmax\ value as a
proxy for mass have also been witnessed in our test cases and we will
come back to it in \Sec{sec:dynamicinfall}.

The first preliminary comparisons focusing on spatial location, \vmax,
and the number of bound particles for the static mock haloes indicate
that even though there exist a variety of different approaches for
halo finding, most of the codes agree with the known correct result
well. If substructure is located close to the centre of the host halo
all the codes tested experienced some difficulties in accurately
recovering it, with all the finders based in 3D configuration space
missing some material. For subhaloes placed near the very centre of
the host halo the more sophisticated 6D finders based in phase space,
while correctly noting the existence of a substructure often
overestimated the associated mass due to confusion with material in
the host halo. After proving to ourselves that we could all
successfully reproduce the location and scale of a supplied mock halo
we performed a resolution study where the mass and hence number of
particles in a subhalo was gradually lowered. We found that
practically all halo finders have a completeness limit of 30-40
particles; substructure objects smaller than this are not reliably
found. Once we had established a firm baseline for our comparisons we
extended the study to consider a full cosmological volume at varying
resolution. The results of this comparison are presented in
\Sec{sec:comparison} below after we first briefly introduce each of
the halo finders involved in the comparison project in \Sec{sec:codes}
and describe the set-up of our mock haloes in \Sec{sec:data}. Finally
we wrap-up and present some conclusions in \Sec{sec:conclusions}.

\section{The Codes} \label{sec:codes}

In this Section we are going to {\em briefly} present the codes that
participated in the halo-finder comparison project. We highlight their
main features allowing for a better understanding of any (possible)
differences in the comparison \Sec{sec:comparison}. The prime
information to be found in each code paragraph should be sufficient to
understand how the algorithm works, how the initial particle content
of a halo is obtained, the way the the (sub-)halo centre and edge are
calculated, how the unbinding is performed and which method of
parallelisation has been applied. Please note that not all halo
finders perform an unbinding, are parallelized or suitable to detect
subhaloes. And we explicitly stress that this Section is neither
intended as a review of all available halo finders nor an elaborate
exposition of the partaking codes; for the latter we refer the reader
to the respective code papers referenced in the subsection of each
halo finder.

As much as possible, the halo finders have been organised in terms of
their methodology: spherical overdensity finders first followed by
FOF-based finders with 6D phase-space finders last. This applies to
both the presentation in this Section as well as the comparison in
\Sec{sec:comparison}.

\subsection{AHF (Knollmann \& Knebe)} \label{sec:ahf}
The MPI+OpenMP parallelised halo finder
\texttt{AHF}\footnote{\texttt{AHF} is freely available from
  \texttt{http://www.popia.ft.uam.es/AMIGA}} \citep[\texttt{AMIGA}
Halo Finder,][]{Knollmann09}, is an improvement of the \texttt{MHF}
halo finder \citep{Gill04a}, which employs a recursively refined grid
to locate local overdensities in the density field. The identified
density peaks are then treated as centres of prospective haloes. The
resulting grid hierarchy is further utilized to generate a halo tree
readily containing the information which halo is a (prospective) host
and subhalo, respectively. We therefore like to stress that our halo
finding algorithm is fully recursive, automatically identifying
haloes, sub-haloes, sub-subhaloes, etc. Halo properties are calculated
based on the list of particles asserted to be gravitationally bound to
the respective density peak. To generate this list of particles we
employ an iterative procedure starting from an initial guess of
particles. This initial guess is based again upon the adaptive grid
hierarchy: for field haloes we start with considering all particles
out to the iso-density contour encompassing the overdensity defined by
the virial criterion based upon the spherical top-hat collapse model;
for subhaloes we gather particles up to the grid level shared with
another prospective (sub-)halo in the halo tree which corresponds to
the upturn point of the density profile due to the embedding within a
(background) host. This tentative particle list is then used in an
iterative procedure to remove unbound particles: In each step of the
iteration, all particles with a velocity exceeding the local escape
velocity, as given by the potential based on the particle list at the
start of the iteration, are removed.  The process is repeated until no
particles are removed anymore. At the end of this procedure we are
left with bona fide haloes defined by their bound particles and we can
calculate their integral and profiled quantities.

The only parameter to be tuned is the refinement criterion used to
generate the grid hierarchy that serves as the basis for the halo tree
and also sets the accuracy with which the centres are being
determined. The virial overdensity criterion applied to find the
(field) halo edges is determined from the cosmological model of the
data though it can readily be tailored to specific needs; for the
analysis presented here we used $200\times\rho_{\rm crit}$. For more
details on the mode of operation and actual functionality we refer the
reader to the two code description papers by \citet{Gill04a} and
\citet{Knollmann09}, respectively.

\subsection{ASOHF (Planelles \& Quilis)} \label{sec:asohf}
The \texttt{ASOHF} finder \citep{Planelles10} is based on the
spherical overdensity (SO) approach. Although it was originally
created to be coupled to an Eulerian cosmological code, in its actual
version, it is a stand-alone halo finder capable of analysing the
outputs from cosmological simulations including different components
(i.e., dark matter, gas, and stars). The algorithm takes advantage of
an AMR scheme to create a hierarchy of nested grids placed at
different levels of refinement. All the grids at a certain level,
named patches, share the same numerical resolution. The higher the
level of refinement the better the numerical resolution, as the size
of the numerical cells gets smaller. The refining criteria are open
and can be chosen depending on the application. For a general purpose,
\texttt{ASOHF} refines when the number of particles per cell exceeds a
user defined parameter. Once the refinement levels are set up, the
algorithm applies the SO method independently at each of those levels.
The parameters needed by the code are the following: i) the
cosmological parameters when analysing cosmological simulations, ii)
the size of the coarse cells, the maximum number of refinement levels
($N_{levels}$), and the maximum number of patches ($N_{patch}$) for
all levels in order to build up the AMR hierarchy of nested grids,
iii) the number of particles per cell in order to choose the cells to
be refined, and iv) the minimum number of particles in a halo.

After this first step, the code naturally produces a tentative list of
haloes of different sizes and masses. Moreover, a complete description
of the substructure (haloes within haloes) is obtained by applying the
same procedure on the different levels of refinement.  A second step,
not using the cells but the particles within each halo, makes a more
accurate study of each of the previously identified haloes.  These
prospective haloes (subhaloes) may include particles which are not
physically bound. In order to remove unbound particles, the local
escape velocity is obtained at the position of each particle. To
compute this velocity we integrate Poisson equation assuming spherical
symmetry.  If the velocity of a particle is higher than the escape
velocity, the particle is assumed to be unbound and is therefore
removed from the halo (subhalo) being considered. Following this
procedure, unbound particles are removed iteratively along a list of
radially ordered particles until no more of them need to be
removed. In the case that the number of remaining particles is less
than a given threshold the halo is dropped from the list.

After this cleaning procedure, all the relevant quantities for the
haloes (subhaloes) as well as their evolutionary merger trees are
computed.  The lists of (bound) particles are used to calculate
canonical properties of haloes (subhaloes) like the position of the
halo centre, which is given by the centre of mass of all the bound
particles, and the size of the haloes, given by the distance of the
farthest bound particle to the centre.

The ability of the \texttt{ASOHF} method to find haloes and their
substructures is limited by the requirement that appropriate
refinements of the computational grid exist with enough resolution to
spot the structure being considered. In comparison to algorithms based
on linking strategies, \texttt{ASOHF} does not require a linking
length to be defined, although at a given level of refinement the size
of the cell can be considered as the linking length of this particular
resolution.

The version of the code used in this comparison is serial, although
there is already a first parallel version based on OpenMP.

\subsection{BDM (Klypin \& Ceverino)} \label{sec:bdm}
The Bound Density Maxima (\texttt{BDM}) halo finder originally
described in \citet{Klypin97} uses a spherical 3D overdensity
algorithm to identify haloes and subhaloes. It starts by finding the
local density at each individual particle position. This density is
defined using a top-hat filter with a constant number of particles
$N_{\rm filter}$, which typically is $N_{\rm filter}=20$. The code
finds all maxima of density, and for each maximum it finds a sphere
containing a given overdensity mass $M_\Delta=(4\pi/3)\Delta\rho_{\rm
  cr}R^3_\Delta$, where $\rho_{\rm cr}$ is the critical density and
$\Delta$ is the specified overdensity.

For the identification of distinct haloes, the code uses the density
maxima as halo centres; amongst overlapping sphere the code finds the
one that has the deepest gravitational potential. Haloes are ranked by
their (preliminary) size and their final radius and mass are derived by
a procedure that guarantees smooth transition of properties of small
haloes when they fall into a larger host) halo becoming subhaloes:
this procedure either assigns $R_\Delta$ or $R_{\rm dist}$ as the
radius for a currently infalling halo as its radius depending on the
environmental conditions, where $R_{\rm dist}$ measures the distance
of the infalling halo to the surface of the soon-to-be host halo.

The identification of subhaloes is a more complicated procedure:
centres of subhaloes are certainly density maxima, but not all density
maxima are centres of subhaloes. \texttt{BDM} eliminates all density
maxima from the list of subhalo candidates which have less than
$N_{\rm filter}$ self-bound particles. For the remaining set of
prospective subhaloes the radii are determined as the minimum of the
following three distances: (a) the distance to the nearest barrier
point (i.e. centres of previously defined (sub-)haloes), (b) the
distance to its most remote bound particle, and (c) the truncation
radius (i.e.  the radius at which the average density of bound
particles has an inflection point). This evaluation involves an
iterative procedure for removing unbound particles and starts with the
largest density maximum.

The unbinding procedure requires the evaluation of the gravitational
potential which is found by first finding the mass in spherical shells
and then by integration of the mass profile. The binning is done in
$\log$ radius with a very small bin size of $\Delta\log(R) =0.005$.

The bulk velocity of either a distinct halo or a subhalo is defined as
the average velocity of the 30 most bound particles of that halo or by
all particles, if the number of particles is less than 30. The number
30 is a compromise between the desire to use only the central
(sub)halo region for the bulk velocity and the noise level.

The code uses a domain decomposition for MPI parallelization and
OpenMP for the parallelization inside each domain.

\subsection{pSO (Sutter \& Ricker)} \label{sec:pSO}
The parallel spherical overdensity (\texttt{pSO}) halo finder is a
fast, highly scalable MPI-parallelized tool directly integrated into
the \texttt{FLASH} simulation code that is designed to provide
on-the-fly halo finding for use in subgrid modeling, merger tree
analysis, and adaptive refinement schemes \citep{Sutter10}. The
\texttt{pSO} algorithm identifies haloes by growing SO spheres. There
are four adjustable parameters, controlling the desired overdensity
criteria for centre detection and halo size, the minimum allowed halo
size, and the resolution of the halo radii relative to the grid
resolution. The algorithm discovers halo centres by mapping dark
matter particles onto the simulation mesh and selecting cell centres
where the cell density is greater than the given overdensity
criterion. The algorithm then determines the halo edge using the SO
radius by collecting particles using the \texttt{FLASH} AMR tree
hierarchy.  The algorithm determines the halo centre, bulk velocity,
mass, and velocity dispersion without additional
post-processing. \texttt{pSO} is provided as both an API for use
in-code and as a stand-alone halo finder.

\subsection{LANL (Luki\'c, Fasel \& Hsu)} \label{sec:LANL}
The \texttt{LANL} halo finder is developed to provide on-the-fly halo
analysis for simulations utilizing hundreds of billions of particles,
and is integrated into the \texttt{MC}$^3$ code \citep{Habib09},
although it can also be used as a stand-alone halo finder. Its core is
a fast $k$D-tree FOF halo finder which uses 3D (block), structured
decomposition to minimize surface to volume ratio of the domain
assigned to each process. As it is aimed at large-scale structure
simulations (100+ Mpc/$h$ on the side), where the size of any single
halo is much smaller than the size of the whole box, it uses the
concept of ``ghost zones'' such that each process gets all the
particles inside its domain as well as those particles which are
around the domain within a given distance (the overload size, a code
parameter chosen to be larger then the size of the biggest halo we
expect in the simulation). After each process runs its serial version
of a FOF finder, MPI based ``halo stitching'' is performed to ensure that
every halo is accounted for, and accounted for only once.

If desired, spherical ``SO'' halo properties can be found using the
FOF haloes as a proxy. Those SO haloes are centred at the particle
with the lowest gravitational potential, while the edge is at
$R_{\Delta}$ -- the radius enclosing an overdensity of $\Delta$. It is
well known that percolation based FOF haloes suffer from the
over-bridging problem; therefore, if we want to ensure completeness of
our SO sample we should run FOF with a smaller linking length than
usual in order to capture all density peaks, but still avoid
over-bridging at the scale of interest (which depends on our choice of
$\Delta$). Overlapping SO haloes are permitted, but the centre of one
halo may not reside inside another SO halo (that would be considered
as a substructure, rather than a ``main'' halo). The physical code
parameters are the linking length for the FOF haloes, and overdensity
parameter $\Delta$ for SO haloes. Technical parameters are the overload
size and the minimum number of particles in a halo.

The \texttt{LANL} halo finder is being included in the standard
distributions of \texttt{PARAVIEW}\footnote{http://www.paraview.org/}
package, enabling researchers to combine analysis and visualization of
their simulations. A substructure finder is currently under
development.

\subsection{SUBFIND (Iannuzzi , Springel \& Dolag)}\label{sec:subfind}
\texttt{SUBFIND} \citep{Springel01subfind} identifies gravitationally
bound, locally overdense regions within an input parent halo,
traditionally provided by a FOF group finder, although other group
finders could be used in principle as well. The densities are
estimated based on the initial set of all particles via adaptive
kernel interpolation based on a number $N_{\rm dens}$ of smoothing
neighbours. For each particle, the nearest $N_{\rm ngb}$ neighbours
are then considered for identifying local overdensities through a
topological approach that searches for saddle points in the isodensity
contours within the global field of the halo. This is done in a
top-down fashion, starting from the particle with the highest
associated density and adding particles with progressively lower
densities in turn. If a particle has only denser neighbours in a
single structure it is added to this region. If it is isolated it
grows a new density peak, and if it has denser neighbours from two
different structures, an isodensity contour that traverses a saddle
point is identified. In the latter case, the two involved structures
are joined and registered as candidate subhaloes if they contain at
least $N_{\rm ngb}$ particles. These candidates, selected according to
the spatial distribution of particles only, are later processed for
gravitational self-boundness. Particles with positive total energy
are iteratively dismissed until only bound particles remain. The
gravitational potential is computed with a tree algorithm, such that
large haloes can be processed efficiently. If the remaining bound
number of particles is at least $N_{\rm ngb}$, the candidate is
ultimately recorded as a subhalo. The set of initial substructure
candidates forms a nested hierarchy that is processed from inside out,
allowing the detection of substructures within substructures. However,
a given particle may only become a member of one substructure,
i.e. \texttt{SUBFIND} decomposes the initial group into a set of
disjoint self-bound structures. Particles not bound to any genuine
substructure are assigned to the ``background halo''. This component
is also checked for self-boundness, so that some particles that are
not bound to any of the structures may remain. For all substructures
as well as the main halo, the particle with the minimum gravitational
potential is adopted as (sub)halo centre. For the main halo,
\texttt{SUBFIND} additionally calculates a SO virial mass around this
centre, taking into account all particles in the simulation (i.e. not
just those in the FOF group that is analyzed). There exist both serial
and MPI-parallelized versions of \texttt{SUBFIND}, which implement the
same underlying algorithms. For more details we refer the reader to
the paper by \citet{Springel01subfind}.

\subsection{FOF (Gottl\"ober \& Turchaninov)}\label{sec:fof}
In order to analyse large cosmological simulations with up to $2048^3$
particles we have developed a new MPI version of the hierarchical
Friends-Of-Friends algorithm with low memory requests. It allows us to
construct very fast clusters of particles at any overdensity
(represented by the linking length) and to deduce the
progenitor-descendant-relationship for clusters in any two different
time steps. The particles in a simulation can consist of different
species (dark matter, gas, stars) of different mass. We consider them
as an undirected graph with positive weights, namely the lengths of
the segments of this graph. For simplicity we assume that all weights
are different. Then one can show that a unique minimum spanning tree
(MST) of the point distribution exists, namely the shortest graph
which connects all points. If subgraphs cover the graph then the MST
of the graph belongs to the union of MSTs of the subgraphs. Thus
subgraphs can be constructed in parallel. Moreover, the geometrical
features of the clusters, namely the fact that they occupy mainly
almost non-overlapping volumes, allow the construction of fast
parallel algorithms.  If the MST has been constructed all possible
clusters at all linking lengths can be easily determined. To represent
the output data we apply topological sorting to the set of clusters
which results in a cluster ordered sequence. Every cluster at any
linking length is a segment of this sequence. It contains the
distances between adjacent clusters. Note, that for the given MST
there exist many cluster ordered sequences which differ in the order
of the clusters but yield the same set of clusters at a desired
linking length.  If the set of particle-clusters has been constructed
further properties (centre of mass, velocity, shape, angular momentum,
orientation etc.) can be directly calculated. Since this concept is by
construction aspherical a circular velocity (as used to characterise
objects found with spherical overdensity algorithms) cannot be
determined here. The progenitor-descendant-relationship is calculated
for the complete set of particles by comparison of the cluster-ordered
sequences at two different output times.

The hierarchical \texttt{FOF} algorithm identifies objects at
different overdensities depending on the chosen linking length
\citep{More11}. In order to avoid artificial misidentifications of
subhaloes on high overdensities one can add an additional
criterion. Here we have chosen the requirement that the spin parameter
of the subhalo should be smaller than one. All subhaloes have been
identified at 512 times the virial overdensity. Thus only the highest
density peak has been taken into account for the mass determination
and the size of the object, which are therefore underestimated. The
velocity of the density peak is estimated correctly but without
removing unbound particles.

\subsection{pFOF (Rasera \& Roy)}\label{sec:pFOF}
Parallel FOF (\texttt{pFOF}) is a MPI-based parallel
Friends-of-Friends halo finder which is used within the DEUS
Consortium \footnote{\texttt{www.deus-consortium.org}} at LUTH
(Laboratory Universe and Theories). It has been parallelized by Roy
and was used for several studies involving large $N$-body simulations
such as \cite{Courtin10, Rasera10}. The principle is the following:
first, particles are distributed in cubic subvolumes of the simulation
and each processor deals with one ``cube'', and runs
Friends-of-Friends locally. Then, if a structure is located close to
the edge of a cube, \texttt{pFOF} checks if there are particles
belonging to the same halo in the neighbouring cube. This process is
done iteratively until all haloes extending across multiple cubes have
been merged. Finally, particles are sorted on a per halo basis, and
the code writes two kinds of output: particles sorted per region,
particles sorted per halo. This makes any post-processing
straightforward because each halo or region can be analysed
individually on a single CPU server. \texttt{pFOF} was successfully
tested on up to 4096 Bluegene/P cores with a $2048^3$ particles
$N$-body simulation. In this article, the serial version was used for
mock haloes and small cosmological simulations, and the parallel
version for larger runs. The linking length was set to $b=0.2$
\citep[however see][for a discussion on the halo
definition]{Courtin10}, and the minimum halo mass to 100
particles. And the halo centres reported here are the centre-of-mass
of the respective particle distribution.

\subsection{Ntropy-fofsv (Gardner, McBride \& Stinson)}\label{sec:ntropy-fofsv}
The Ntropy parallel programming framework is derived from N-body codes
to help address a broad range of astrophysical problems
\footnote{http://www.phys.washington.edu/users/gardnerj/ntropy}.  This
includes an implementation of a simple but efficient FOF halo finder,
\texttt{Ntropy-fofsv}, which is more fully described in
\citet{Gardner07a} and \citet{Gardner07b}. Ntropy provides a
``distributed shared memory'' (DSM) implementation of a $k$D-tree,
where the application developer can reference tree nodes as if they
exist in a global address space, even though they are physically
distributed across many compute nodes.  Ntropy uses the $k$D-tree data
structures to speed up the FOF distance searches.  It also employs an
implementation of the \citet{Shiloach82} parallel connectivity
algorithm to link together the haloes that span separate processor
domains.  The advantage of this method is that no single computer node
requires knowledge of all of the groups in the simulation volume,
meaning that \texttt{Ntropy-fofsv} is scalable to petascale platforms
and handle large data input. This algorithm was used in the mock halo
test cases to stitch together particle groups found across many
threads into the one main FOF halo. As FOF is a deterministic
algorithm, \texttt{Ntropy-fofsv} takes a single physical linking
length to group particles into FOF haloes without any performing
particle unbinding or subhalo identification.  The halo centres for
the analysis presented here use centre-of-mass estimates based on the
FOF particle list.  Ntropy achieves parallelisation by calling
``machine dependent library'' (MDL) that consists of high-level
operations such as ``acquire\_treenode'' or ``acquire\_particle.''  This
library is rewritten for a variety of models (MPI, POSIX Threads, Cray
SHMEM, etc.), allowing the framework to extract the best performance
from any parallel architecture on which it is run.

\subsection{VOBOZ (Neyrinck)} \label{sec:VOBOZ}
Conceptually, a \texttt{VOBOZ} \citep[VOronoi BOund
  Zones,][]{Neyrinck05} halo or subhalo is a density peak surrounded
by gravitationally bound particles that are down steepest-density
gradients from the peak.  A statistical significance is measured for
each (sub)halo, based on the probability that Poisson noise would
produce it.

The only physical parameter in \texttt{VOBOZ} is the density threshold
characterizing the edge of (parent) haloes (set to 200 times the mean
density here), which typically only affects their measured masses.  To
return a definite halo catalog, we also impose a
statistical-significance threshold (set to 4-$\sigma$ here), although
depending on the goal of a study, this may not be necessary.

Density peaks are found using a Voronoi tessellation (parallelizable
by splitting up the volume), which gives an adaptive, parameter-free
estimate of each particle's density and set of neighbours
\citep[e.g.][]{Schaap00}.
Each particle is joined to the peak particle
(whose position is returned as the halo centre) that lies up the
steepest density gradient from that particle.  A halo associated with
a high density peak will also contain smaller density peaks.  The
significance of a halo is judged according to the ratio of its central
density to a saddle point joining the halo to a halo with a higher
central density, comparing to a Poisson point process.  Pre-unbinding
(sub)halo boundaries are defined along these density ridges.

Unbinding evaporates many spurious haloes, and often brings other halo
boundaries inward a bit, reducing the dependence on the outer density
contrast.  Particles not gravitationally bound to each halo are
removed iteratively, by comparing their potential energies (measured
as sums over all other particles) to kinetic energies with respect to
the velocity centroid of the halo's core (i.e.\ the particles that
directly jump up density gradients to the peak).  The unbinding is
parallelized using OpenMP.  In the cosmological test, we remove haloes
with fewer than 20 particles from the \texttt{VOBOZ} halo list.

\subsection{ORIGAMI (Falck, Neyrinck \& Aragon-Calvo)} \label{sec:ORIGAMI}
\texttt{ORIGAMI} \citep[Order-ReversIng Gravity, Apprehended Mangling
Indices,][]{Falck11} uses a natural, parameter-free definition of the
boundary between haloes and the non-halo environment around them: halo
particles are particles that have experienced shell-crossing.  This
dynamical definition does not make use of the density field, in which
the boundary can be quite ambiguous.  In one dimension, shell
crossings can be detected by looking for pairs of particles whose
positions are out-of-order compared with their initial positions. In
3D, then, a halo particle is defined as a particle that has undergone
shell crossings along 3 orthogonal axes. Similarly, this would be 2
axes for a filament, 1 for a wall, and 0 for a void. There is a huge
number of possible sets of orthogonal axes in the initial grid to use
to test for shell-crossing, but we only used four simple ones, which
typically suffice to catch all the shell-crossings.  We used the
Cartesian $x$, $y$, and $z$ axes, as well as the three sets of axes
consisting of one Cartesian axis and two ($45\degr$) diagonal axes in
the plane perpendicular to it.

Once halo particles have been tagged, there are many possible ways of
grouping them into haloes.  For this paper, we grouped them on a
Voronoi tessellation of final-conditions particle positions.  This
gives a natural density estimate \citep[e.g.][\texttt{VTFE}, Voronoi
Tessellation Field Estimator]{Schaap00} and set of neighbours for each
particle.  Haloes are sets of halo particles connected to each other
on the Voronoi tessellation.  To prevent haloes from being unduly
linked, we additionally require that a halo contain at most one halo
``core'', defined as a set of particles connected on the tessellation
that all exceed a \texttt{VTFE} density threshold. This density
threshold is the only parameter in our algorithm, since the initial
tagging of halo particles is parameter-free; for this study, we set it
to 200 times the mean density.  We partition connected groups of halo
particles with multiple cores into haloes as follows: each core
iteratively collects particles in concentric rings of Voronoi
neighbours until all halo particles are associated.  The tagging
procedure establishes halo boundaries, so no unbinding procedure is
necessary.  Also, we note that currently, the algorithm does not
identify subhaloes. We remove haloes with fewer than 20 particles from
the \texttt{ORIGAMI} halo catalogue, and the halo centre reported is
the position of the halo's highest-density particle.

\textit{Please note that due to its nature ORIGAMI is only applicable
  to cosmological simulations and hence only enters the comparison
  project in the respective Section~\ref{sec:cosmologicalsimulation}.}

\subsection{SKID (Stadel \& Potter)} \label{sec:SKID}
\texttt{SKID} (Spline Kernel Interpolative Denmax)\footnote{The OpenMP
  parallelized version of \texttt{SKID} can be freely downloaded from
  \texttt{http://www.hpcforge.org}}, first mentioned in
\citet{Governato97} and extensively described in \citet{Stadel01},
finds density peaks within $N$-body simulations and subsequently
determines all associated bound particles thereby identifying
haloes. It is important to stress that \texttt{SKID} will only find
the smallest scale haloes within a hierarchy of haloes as is generally
seen in cosmological structure formation simulations.  Unlike original
DENMAX \citep{Bertschinger91, Gelb92} which used a fixed grid based
density estimator, \texttt{SKID} uses SPH (i.e., smoothed particle
hydrodynamics) kernel averaged densities which are much better suited
to the Lagrangian nature of $N$-body simulations and allow the method
to locally adapt to the large dynamic range found in cosmological
simulations.

Particles are slowly slid (each step moving the particles by a
distance of order the softening length in the simulation) along the
local density gradient until they pool at a maximum, each pool
corresponding to each initial group. This first phase of \texttt{SKID}
can be computationally very expensive for large simulations, but is
also quite robust.

Each pool is then ``unbound'' by iteratively evaluating the binding
energy of every particle in their original positions and then removing
the most non-bound particle until only bound particles remain.  This
removes all particles that are not part of substructure either because
they are part of larger scale structure or because they are part of
the background.

\texttt{SKID} can also identify structure composed of gas and stars in
hydrodynamical simulations using the dark matter only for its
gravitational binding effect. The ``Haloes going MAD'' meeting has
motivated development of an improved version of the algorithm capable
of also running on parallel computers.

\subsection{AdaptaHOP (Tweed \& Colombi)} \label{sec:adaptahop}
The code \texttt{AdaptaHOP} is described in Appendix A of
\citet{Aubert04}. The first step is to compute an SPH density for each
particle from the 20 closest neighbours. Isolated haloes are then
described as groups of particles above a density threshold $\rho_t$,
where this parameter is set to 80, which closely matches results of a
FOF group finder with parameter $b=0.2$. To identify subhaloes within
those groups, local density maxima and saddle points are
detected. Then, by increasing the density threshold, it is a simple
matter to decompose haloes into nodes that are either density maxima,
or groups of particles whose density is between two values of saddle
points. A node structure tree is then created to detail the whole
structure of the halo itself. Each leaf of this tree is a local
density maximum and can be interpreted as a subhalo. However, further
post-processing is needed to define the halo structure tree,
describing the host halo itself, its subhaloes and subhaloes within
subhaloes. This part of the code is detailed in \citet{Tweed09}; the
halo structure tree is constructed so that the halo itself contains
the most massive local maximum (Most massive Sub maxima Method: MSM).
This method gives the best result for isolated snapshots, as used in
this paper.

In more detail, \texttt{AdaptaHOP} needs a set of seven
parameters. The first parameter is the number of neighbours $n_{nei}$
used with a $k$D-tree scheme in order to estimate the SPH
density. Among these $n_{nei}$ neighbours, the $n_{hop}$ closest are
used to sweep through the density field and detect both density maxima
and saddle points. As previously mentioned, the parameter $\rho_t$
sets the halo boundary. The decomposition of the halo itself into
leaves that are to be redefined as subhaloes has to fulfil certain
criteria set by the remaining four parameters. The most relevant is
the statistical significance threshold, set via the parameter $fudge$,
defined via $(\langle\rho\rangle - \rho_t)/\rho_t > fudge/\sqrt{N}$,
where $N$ is the number of particles in the leaves. The minimal mass
of a halo is limited by the parameter $n_{members}$, the minimum
number of particles in a halo.  Any potential subhalo has also to
respect two conditions with respect to the density profile and the
minimal radius, through the parameters $\alpha$ and
$f_{\epsilon}$. These two values ensure that a subhalo has a maximal
density $\rho_{max}$ such as $\rho_{max} > \alpha\langle\rho\rangle$
and a radius greater than $f_{\epsilon}$ times the mean interparticle
distance. We used the following set of parameters
($n_{nei}=n_{hop}=20$, $\rho_t=80$, $fudge=4$, $\alpha=1$,
$f_{\epsilon}=0.05$, $n_{members}=20$).  It is important to understand
that all nodes are treated as leaves and must comply with
aforementioned criteria before being further decomposed into separate
structures. As for defining haloes and subhaloes themselves, this is
done by grouping linked lists of particles corresponding to different
nodes and leaves from the node structure tree. Further, the halo and
subhalo centres are defined as the position of the particle with the
highest density. The halo edge corresponds to the $\rho_t$ density
threshold, whereas the saddle points define the subhalo edge.

Please note that \texttt{AdaptaHOP} is a mere topological code that
does \textit{not} feature an unbinding procedure. For substructures
(whose boundaries are chosen from the saddle point value) this may
impact on the estimate of the mass as well as lead to contamination by
host particles.

\subsection{HOT (Ascasibar)} \label{sec:hot}
This algorithm, still under development, computes the Hierarchical
Overdensity Tree, (\texttt{HOT}), of a point distribution in an arbitrary
multidimensional space.  \texttt{HOT} is introduced as an alternative to
the minimal spanning tree (MST) for spaces where a metric is not well
defined, like the phase space of particle positions and velocities.

The method is based on the Field Estimator for Arbitrary Spaces
\citep[FiEstAS, ][]{Ascasibar05, Ascasibar10}.  First, the space
is tessellated one dimension at a time, until it is divided into a set
of hypercubical cells containing exactly one particle.  Particles in
adjacent cells are considered as neighbours.  Then, the mass of each
point is distributed over an adaptive smoothing kernel as described in
\citet{Ascasibar10}, which provides a key step in order to define a
metric.

In the \texttt{HOT+FiEstAS} scheme, objects correspond to the peaks of
the density field, and their boundaries are set by the isodensity
contours at the saddle points.  At each saddle point, the object
containing less particles is attached to the most massive one, which
may then be incorporated into even more massive objects in the
hierarchy.  This idea can be implemented by computing the MST of the
data distribution, defining the distance between two neighbouring
particles as the minimum density along an edge connecting them
(i.e. the smallest of the two densities, or the density of the saddle
point when it exists).  However, this is not practical for two
reasons.  Firstly, defining a path between two particles is not
trivial when a metric is not available.  Secondly, finding the saddle
points would require a minimisation along the path, which is extremely
time consuming when a large number of particles is involved.  These
problems may be overcome if the distance between two data points is
given by the average density within the hyperbox they define.

Once the distances are defined in this way, \texttt{HOT+FiEstAS}
computes the MST of the data distribution by means of Kruskal's
algorithm \citep{Kruskal56}.  The output of the algorithm consists of
the tree structure, given by the parent of each data point in
\texttt{HOT}, and a catalogue containing an estimate of the centroid
(given by the density-weighted centre of mass) as well as the number
of particles in the object (both including and excluding
substructures).  In order to discard spurious density fluctuations, a
minimum number of points and density contrast are required for an
object to be output to the catalogue.  Currently, these parameters are
set to $N>20$ particles and a contrast threshold $\rho_{\rm
peak}/\rho_{\rm background}>5$.  Although these values seem to yield
reasonable results, more experimentation is clearly needed.

In this work, the algorithm is applied to the particle positions only
(\texttt{HOT3D}) as well as the full set of phase-space coordinates
(\texttt{HOT6D}).  Since it is intended as a general data analysis
tool, not particularly optimised for the problem of halo
identification, it should not (and does not) take into account any
problem-specific knowledge such as the concepts of binding energy or
virial radius. The latter quantity, as well as the maximum circular
velocity, have been computed from the raw particle IDs returned by the
code.

The definition of object boundaries in terms of the saddle points of
the density field will have a relatively mild impact in the results
concerning the mock haloes, but it is extremely important in the
cosmological case. \texttt{HOT}+\texttt{FiEstAS} will, for instance,
identify large-scale filamentary structures that are not considered
haloes by any of the other algorithms (although many of these objects are
indeed gravitationally bound).

On the other hand, keeping unbound particles will be an issue for
subhaloes close to the centre of their host, especially in three
dimensions, and a post-processing\footnote{\texttt{HOT3D} does not
  even read particle velocities} script will be developed to perform
this task.

\textit{Please note that due to its present implementation \texttt{HOT} is not yet
  applicable to cosmological simulations and hence only enters the
  comparison project in the mock halo Section~\ref{sec:mockhaloes}.}

\subsection{HSF (Maciejewski)} \label{sec:HSF}
The Hierarchical Structure Finder \citep[\texttt{HSF},][]{Maciejewski09}
identifies objects as connected self-bound particle sets above some
density threshold.  This method consists of two steps. Each particle
is first linked to a local DM phase-space density maximum by following
the gradient of a particle-based estimate of the underlying DM
phase-space density field. The particle set attached to a given
maximum defines a candidate structure. In a second step, particles
which are gravitationally unbound to the structure are discarded until
a fully self-bound final object is obtained.

In the initial step the phase-space density and phase-space gradients
are estimated by using a six-dimensional SPH smoothing kernel with a
local adaptive metric as implemented in the \texttt{EnBiD} code
\citep{Sharma06}. For the SPH kernel we use $N_{\rm sph}$ between $20$
and $64$ neighbours whereas for the gradient estimate we use $N_{\rm
  ngb}=20$ neighbours.

Once phase-space densities have been calculated, we sort the particles
according to their density in descending order. Then we start to grow
structures from high to low phase-space densities. While walking down
in density we mark for each particle the two closest (according to the
local phase-space metric) neighbours with higher phase-space density,
if such particles exist. In this way we grow disjoint structures until
we encounter a saddle point, which can be identified by observing the
two marked particles and seeing if they belong to different
structures. A saddle point occurs at the border of two
structures. According to each structure mass, all the particles below
this saddle point can be attached to only one of the structures if it
is significantly more massive than the other one, or redistributed
between both structures if they have comparable masses.  This is
controlled by a simple but robust cut or grow criterion depending on a
{\em connectivity parameter} $\alpha$ which is ranging from $0.2$ up
to $1.0$. In addition, we test on each saddle point if structures are
statistically significant when compared to Poisson noise (controlled by
a $\beta$ parameter). At the end of this process, we obtain a
hierarchical tree of structures.

In the last step we check each structure against an unbinding
criterion. Once we have marked its more massive partner for each
structure, we sort them recursively such that the larger partners
(parents) are always after the smaller ones (children). Then we
unbind structure after structure from children to parents and add
unbound particles to the larger partner. If the structure has less
than $N_{\rm cut}=20$ particles after the unbinding process, then we
mark it as not bound and attach all its particles to its more
massive partner (note, that a smaller $N_{\rm cut}$ is used for the
resolution study in \Sec{sec:resolution}). The most bound 
particle of each halo/subhalo defines its position centre.

Although \texttt{HSF} can be used on the entire volume, to speed up the
process of identification of the structures in the cosmological
simulation volume we first apply the FOF method to disjoint the
particles into smaller FOF groups.

\subsection{6DFOF (Zemp \& Diemand)} \label{sec:6dfof}
\texttt{6DFOF} is a simple extension of the well known FOF method
which also includes a proximity condition in velocity space.  Since
the centres of all resolved haloes and subhaloes reach a similar peak
phase space density they can all be found at once with \texttt{6DFOF}.
The algorithm was first presented in \cite{Diemand06}.  The
\texttt{6DFOF} algorithm links two particles if the following
condition
\begin{equation}
\frac{({\bf x}_1 - {\bf x}_2)^2}{\Delta x^2} +
\frac{({\bf v}_1 - {\bf v}_2)^2}{\Delta v^2} < 1
\end{equation}
is fulfilled.  There are three free parameters: $\Delta x$, the
linking length in position space, $\Delta v$, the linking length in
velocity space, and $N_\mathrm{min}$, the minimum number of particles
in a linked group so that it will be accepted.  For $\Delta v
\rightarrow \infty$ it reduces to the standard FOF scheme.  The
\texttt{6DFOF} algorithm is used for finding the phase space
coordinates of the high phase space density cores of haloes on all
levels of the hierarchy and is fully integrated in parallel within the
MPI and OpenMP parallelised code \texttt{PKDGRAV} \citep{Stadel01}.

The centre position and velocity of a halo are then determined from
the linked particles of that halo.  For the centre position of a halo,
one can choose between the following three types: 1) the
centre-of-mass of its linked particles, 2) the position of the
particle with the largest absolute value of the potential among its
linked particles or 3) the position of the particle which has the
largest local mass density among its linked particles.  For the
analysis presented here, we chose type 3) as our halo centre position
definition.  The centre velocity of a halo is calculated as the
centre-of-mass velocity of its linked particles.  Since in
\texttt{6DFOF} only the particles with a high phase space density in
the very centre of each halo (or subhalo) are linked together, it
explains the somewhat different halo velocities (compared to the other
halo finders) and slightly offset centres in cases only a few
particles were linked.

Other properties of interest (e.g. mass, size or maximum of the
circular velocity curve) and the hierarchy level of the individual
haloes are then determined by a separate profiling routine in a post
processing step.  For example, a characteristic size and mass scale
definition (e.g. $r_\mathrm{200c}$ and $M_\mathrm{200c}$) for field
haloes based on traditional spherical overdensity criteria can be
specified by the user.  For subhaloes, a truncation scale can be
estimated as the location where the mass density profile reaches a
user specified slope.  During the profiling step no unbinding
procedure is performed.  Hence, the profiling step does not base its
(sub-)halo properties upon particle lists but rather on spherical
density profiles.  Therefore, \texttt{6DFOF} directly returned halo
properties instead of the (requested) particle ID lists.

\subsection{Rockstar (Behroozi)} \label{sec:rockstar}
\texttt{Rockstar} is a new phase-space based halo finder designed to
maximize halo consistency across timesteps; as such, it is especially
useful for studying merger trees and halo evolution (Behroozi et
al. in prep.).  \texttt{Rockstar} first selects particle groups with a
3D Friends-of-Friends variant with a very large linking length
($b=0.28$).  For each main FOF group, \texttt{Rockstar} builds a
hierarchy of FOF subgroups in phase space by progressively and
adaptively reducing the linking length, so that a tunable fraction
(70\%, for this analysis) of particles are captured at each subgroup
as compared to the immediate parent group.  For each subgroup, the
phase-space metric is renormalized by the standard deviations of
particle position and velocity.  That is, for two particles $p_1$ and
$p_2$ in a given subgroup, the distance metric is defined as:
\begin{equation}
d(p_1, p_2) = \left(\frac{({\bf x}_1-{\bf x}_2)^2}{\sigma_x^2} +
  \frac{({\bf v}_1-{\bf v}_2)^2}{\sigma_v^2}\right)^{1/2},
\end{equation}
where $\sigma_x$ and $\sigma_v$ are the particle position and velocity
dispersions for the given subgroup. This metric ensures an adaptive
selection of overdensities at each successive level of the FOF
hierarchy.

When this is complete, \texttt{Rockstar} converts FOF subgroups into
haloes beginning at the deepest level of the hierarchy.  For a subgroup
without any further sublevels, all the particles are assigned to a
single seed halo.  If the parent group has no other subgroups, then
all the particles in the parent group are assigned to the same seed
halo as the subgroup.  However, if the parent group has multiple
subgroups, then particles are assigned to the subgroups' seed haloes
based on their phase-space proximity.  In this case, the phase-space
metric is set by halo properties, so that the distance between a halo
$h$ and a particle $p$ is defined as:
\begin{equation}
d(h, p) = \left(\frac{({\bf x}_h-{\bf x}_p)^2}{r_{vir}^2} +
  \frac{({\bf v}_h-{\bf v}_p)^2}{\sigma_v^2}\right)^{1/2},
\end{equation}
where $r_{vir}$ is the current virial radius of the seed halo and
$\sigma_v$ is the current particle velocity dispersion.  This process
is repeated at all levels of the hierarchy until all particles in the
base FOF group have been assigned to haloes.  Unbinding is performed
using the full particle potentials (calculated using a modified Barnes
\& Hut method, \citet{Barnes86}); halo centres are defined by
averaging particle positions at the FOF hierarchy level which yields
the minimum estimated Poisson error---which in practice amounts to
averaging positions in a small region close to the phase-space density
peak.  For further details about the unbinding process and for details
about accurate calculation of halo properties, please see Behroozi et
al. in prep.

\texttt{Rockstar} is a massively parallel code (hybrid OpenMP/MPI
style); it can already run on up to $10^5$ CPUs and on the very
largest simulations ($> 10^{10}$ particles).  Additionally, it is very
efficient, requiring only 56 bytes of memory per particle and 4-8
(total) CPU hours per billion particles in a simulation snapshot. The
code is in the final stages of development; as such, the results in
this paper are a minimum threshold for the performance and accuracy of
the final version.\footnote{Those interested in obtaining a copy of
  the code as well as a draft of the paper should contact the author
  at behroozi@stanford.edu.  Current acceptable input formats for
  simulation files are \texttt{ART}, \texttt{GADGET-2}, and ASCII.}

\section{The Data} \label{sec:data}
In order to study, quantify, and assess the differences between
various halo finding techniques we first have to define a unique set
of test cases. In that regard we decided to split the suite of
comparisons into two major parts: 

\begin{itemize}
 \item a well-defined mock haloes
consisting of field haloes in isolation as well as (sub-)subhaloes
embedded within the density background of larger entities, and
 \item a state-of-the-art cosmological simulation primarily focusing on
  the large-scale structure.
\end{itemize}

We further restricted ourselves to analysing dark matter only data
sets as the inclusion of baryons (especially gas and its additional
physics) will most certainly complicate the issue of halo finding. As
most of the codes participating in this comparison project do not
consider gas physics in the process of object identification we
settled for postponing such a comparison to a later study.

We further adopted the following strategy for the comparison.
For the mock haloes each code was asked to return a list of particles
and the centre of the (sub-)halo as derived from applying the halo
finder to the respective data set. These centres and particle lists
were then post-processed by one single code deriving all the
quantities studied below. By this approach we aimed at homogenising
the comparison and eliminating subtle code-to-code variations during
the analysis process. However, we also need to acknowledge that not
all codes complied with this request as they were not designed to
return particle lists; those codes nevertheless provided the halo
properties in question and are included in the comparison.

For the comparison of the cosmological simulations each code merely
had to return those halo properties to be studied, based upon each
and every code individually. The idea was to compare the actual
performance of the codes in a realistic set-up without interference in
the identification/analysis process.

\subsection{Mock Haloes} \label{sec:mockdata}
In order to be able to best quantify any differences in the results
returned by the different halo finders it is best to construct test
scenarios for which the correct answer is known in advance. Even
though we primarily aim at comparing \vmax\ and the number of
gravitationally bound particles we also want to have full control over
various definitions of, for instance, virial mass, i.e. we require
haloes whose density profile is well known. Additionally, as subhalo
detection is of prime interest in state-of-the-art cosmological
simulations we also place haloes within haloes within haloes
etc. Further, sampling a given density profile with particles also
gives us the flexibility to study resolution effects related to the
number of particles actually used.

We primarily used the functional form for the (dark matter) density
profile of haloes originally proposed in a series of papers by
Navarro, Frenk \& White \citep[][the so-called ``NFW
profile"]{Navarro95, Navarro96, Navarro97},
\begin{equation}
 \frac{\rho(r)}{\rho_{\rm crit}}=\frac{\delta_{\rm c}}{r/r_{\rm s}(1+r/r_{\rm s})^2},
\end{equation}
\noindent where $\rho_{\rm crit}$ is the critical density of the universe,
$r_{\rm s}$ is the scale radius and $\delta_{\rm c}$ is the characteristic
density.  NFW haloes are characterised by their mass for a given
enclosed overdensity,
\begin{equation}
 M_{\Delta}=\frac{4\pi}{3}r^3_{\Delta}\Delta\rho_{\rm crit},
\end{equation}
\noindent where $\Delta$ is a multiple of the critical density that defines the
magnitude of the overdensity and $r_{\Delta}$ is the radius at which this occurs.
The characteristic density is then defined as,
\begin{equation}
 \delta_{\rm c} =\frac{\Delta}{3}\frac{c^3}{\ln(1+c)-c/(1+c)},
\end{equation}
\noindent where $c = r_{\Delta}/r_{\rm s}$ is the concentration. The
mock haloes were generated with using a predefined number of particles
that reproduced the NFW profile even though the consensus has moved
away from the statement that dark matter haloes follow this particular
profile all the way down to the centre.  We are not interested in
probing those very central regions where the density profile starts to
deviate from the NFW form as found nowadays in cosmological
simulations \citep{Stadel09, Navarro10}. We need to stress that the
position and size of the maximum of the rotation curve is in fact
unaffected in all tests presented here.  The velocities of the
particles were then assigned using the velocity dispersion given in
\citet{Lokas01} and distributed using a Maxwell-Boltzmann function
\citep{Hernquist93}.\footnote{We are aware that the velocity
  distribution is not derived from the full distribution function and
  that the Maxwell-Boltzmann distribution is only an approximation
  \citep[cf.][]{Kazantzidis04c, Zemp08}. Despite this, it will have no
  effect on the ability of halo finders to recover the haloes as has
  been shown in \citet{Muldrew11} where also more details about the
  generation of the mock haloes can be found.}

In addition to mock haloes whose density profile is based upon the
findings in cosmological simulations (at least down to those scales
probed here) we also chose to generate test haloes that follow a
Plummer profile \citep{Plummer11},
\begin{equation}
 \rho(r)=\frac{3M}{4\pi r^3_{\rm s}}(1+r^2/r^2_{\rm s})^{-\frac{5}{2}},
\end{equation}
\noindent where $M$ is the total mass and $r_{\rm s}$ is the scale radius. The
mock haloes were then produced again using a predefined number of particles to
reproduce the profile, but this time the velocities were obtained using an
isotropic, spherically symmetric distribution function \citep{Binney87}.
The two major differences between the Plummer and the NFW density profile are
that for the former profile the mass converges and it contains a well defined
constant-density core. This constant density may pose problems for halo
finders as most of them rely on identifying peaks in the density field as
(potential) sites for dark matter haloes. We stress that the Plummer
spheres are intended as academic problems with no observed counter-part in
cosmological simulations and hence only to be taken lightly and for
information purposes; they may be viewed as a stability test for halo
finders and as a trial how sensitive halo characteristics are against
precise measurements of the centre. We will see that some properties
can still be stably recovered even if an incorrect determination of
the Plummer halo centre is made. 

As we also plan to study the accurate recovery of substructure we
generated setups where one (or multiple) subhaloes are embedded within
the density profile of a larger host halo. To this end we generate,
for instance, two haloes in isolation: one of them (the more massive
one) will then serve as the host whereas the lighter one will be
placed inside at a known distance to the centre of its host and with a
certain (bulk) velocity. The concentrations (i.e. the ratio between
the virial and the scale radius) have been chosen in order to meet the
findings of cosmological simulations \citep[e.g.][]{Bullock01}. All
our mock haloes are set-up with fully sampled 6D initial phase space
distributions and every halo (irrespective of it becoming a host or a
subhalo) has been evolved in isolation for several Gyrs in order to
guarantee equilibrium. The mass of all particles in both the host halo
and the subhalo are identical and all haloes have been sampled with
particles out to $2\times R_{100}$ where $R_{100}$ marks the point
where the density drops below $\rm {100}\times\rho_{\rm crit}$. For
more details of the procedure and the generation of the NFW haloes we
would like to refer the reader to \citet{Muldrew11} and for the
generation of the Plummer spheres to \citet{Read06}.

The characteristics of the haloes are summarised in
\Tab{tab:static}. We are aware of the fact that even though the
radius at which the enclosed overdensity reaches some defined level 
is well-defined for our subhaloes when they were generated in
isolation, such a definition becomes obsolete once they are placed
inside a host. However, we nevertheless need to acknowledge that such
a definition may serve as a fair basis for the comparisons of the
recovery of subhalo properties amongst different halo finders.

Further, placing an unmodified subhalo at an arbitrary radial distance
within a parent halo is also in part an academic exercise. It neglects
that ``real'' subhaloes will always be tidally truncated. In that
regards, it is not realistic to have an extended/untruncated subhalo
at small distances to the host's centre. Some halo finders
(e.g. \texttt{SUBFIND}) rely on the tidal truncation in order to be
able to avoid a very large radially dependent bias in the amount of
mass that can be recovered for a subhalo.\\

\noindent
For each of the two types of density profile we generated the following setups:

\begin{enumerate}
 \item isolated host halo
\\
 \item isolated host halo + subhalo at $0.5 R_{\rm 100}^{\rm host}$
\\
 \item isolated host halo + subhalo at $0.5 R_{\rm 100}^{\rm host}$

                                       \hspace*{2.8cm} + subsubhalo
                                       at ($0.5 R_{\rm 100}^{\rm host}+0.5 R_{\rm 100}^{\rm subhalo}$)
\\
\item isolated host halo + 5 subhaloes at various distances
\\
\end{enumerate}

\noindent
The (sub-)subhaloes were placed along the $x$-axis and given radially
infalling bulk velocities of 1000 km/sec for the subhalo and 1200
km/sec for the subsubhalo, respectively. These velocities are typical
for what you would expect in a dark matter host halo and were set to
round numbers to make the analysis easier; their values were
motivated by $\sqrt{2GM_{\rm host}(<D)/D}$ where $D$ is the distance
of the subhalo to the host's centre.

The first three setups were used to study the overall recovery of
(sub-)halo properties presented in \Sec{sec:static}. The fourth test
has been used to study the radial dependence of subhalo properties
introduced in \Sec{sec:radialdependence}.

Besides of the recovery of (sub-)halo properties we also aim at
answering the question ``How many particles are required to find a
subhalo?". To this end we systematically lowered the number of
particles (and hence also the subhalo mass as our particle mass
remains constant) used to sample the subhalo listed above as test case
\#2.  The properties of these mock subhaloes are summarised in
\Tab{tab:resolution} and the results will be shown in
\Sec{sec:resolution}.

\begin{table*}
  \caption{The properties of the (sub-)haloes for the study of
    recovered halo properties presented in \Sec{sec:static} and
    \Sec{sec:radialdependence} . The number of particles $N_{\rm xxx}$
    counts all particles out to $R_{\rm xxx}$ where the density drops
    below $\rm {xxx}\times\rho_{\rm crit}$. Masses are given in
    \hMsun, radii in \hkpc, and velocities in km/sec. Please note that
    all haloes have been sampled out to $2\times R_{\rm 100}$ and that
    the Plummer subsubhalo does not reach this overdensity and has
    been truncated at 23.9\hkpc. The halo type indicates whether the
    halo is a host, a subhalo or a subsubhalo.  $R_{\rm s}$ is the
    scale length of the appropriate halo type.  }
\label{tab:static}
\begin{center}
\begin{tabular}{llllllllll}
\hline
profile    & type & $N_{\rm 100}$ & $M_{\rm 100}$ & $R_{\rm 100}$ & $N_{\rm 200}$ & $M_{\rm 200}$ & $R_{\rm 200}$ & $R_{\rm s}$ & \vmax\\
\hline
NFW            & host      & 10$^6$ & 10$^{14}$ & 947.4 & 760892 & 7.61 $\times 10^{13}$ & 689.1 & 189.5 & 715\\
               & sub       & 10$^4$ & 10$^{12}$ & 204.1 &  8066  & 8.07 $\times 10^{11}$ & 151.4 & 17.0  & 182\\
               & subsub    & 10$^2$ & 10$^{10}$ & 44.0  &  84    & 8.42 $\times 10^{9}$  &  33.1 & 2.6   & 43\\
Plummer        & host      & 10$^6$ & 10$^{14}$ & 947.0 & 966326 & 9.66 $\times 10^{13}$ & 760.5 & 190.0 & 961\\
               & sub       & 10$^4$ & 10$^{12}$ & 204.0 &  9937  & 9.94 $\times 10^{11}$ & 161.7 & 17.0  & 314\\
               & subsub    & 10$^2$ & 10$^{10}$ & 23.9  &  100   & 10.00$\times 10^{9}$  & 23.9  & 2.6   & 79\\
\hline
\end{tabular}
\end{center}
\end{table*}

\begin{table}
  \caption{The properties of the subhaloes for the NFW resolution study
    presented in \Sec{sec:resolution}. Radii are given in \hkpc, and velocities in km/sec.}
\label{tab:resolution}
\begin{center}
\begin{tabular}{lllll}
\hline
$N_{\rm 100}$ & $N_{\rm tot}$ & $R_{\rm 100}$ & \vmax & $R_{\rm vmax}$\\
\hline
10 & 13 & 20.41 & 18.24 & 3.68\\
20 & 27 & 25.72 & 22.99 & 4.62\\
30 & 41 & 29.44 & 26.31 & 5.30\\
40 & 55 & 32.40 & 28.96 & 5.85\\
50 & 68 & 34.90 & 31.20 & 6.30\\
100 & 137 & 43.98 & 39.31 & 7.93\\
500 & 687 & 75.20 & 67.21 & 13.55\\
1000 & 1375 & 94.74 & 84.68 & 17.08\\
\hline
\end{tabular}
\end{center}
\end{table}

Besides these well controlled tests we also performed a so-called
``blind test'' where the precise set-up of the data to be analysed by
each halo finder was unknown to the participants. We introduce this
particular experiment alongside its results in a stand-alone
Section~\ref{sec:blindtest}. Only a small subset of the halo finders
took part in this trial.

We close this section with a cautionary remark that not all halo
finders are ab initio capable of identifying subhaloes and hence some
of the test cases outlined here were not performed by all the
finders. Therefore some of the codes only contribute data points for
the host halo in \Sec{sec:comparison}.

\subsection{Cosmological Simulation} \label{sec:cosmodata}
The cosmological simulation used for the halo-finder code comparison
project is the so-called MareNostrum Universe which was performed with
the entropy conserving GADGET2 code \citep{Springel05}. It followed
the nonlinear evolution of structures in gas and dark matter from
$z=40$ to the present epoch ($z=0$) within a comoving cube of side
500\hMpc. It assumed the spatially flat concordance cosmological model
with the following parameters: the total matter density
$\Omega_m=0.3$, the baryon density $\Omega_b=0.045$, the cosmological
constant $\Omega_{\Lambda}=0.7$, the Hubble parameter $h=0.7$, the
slope of the initial power spectrum $n=1$, and the normalisation
$\sigma_8=0.9$. Both components, the gas and the dark matter, were
resolved by $1024^3$ particles, which resulted in a mass of $m_{\rm
  DM}=8.3\times 10^{9}$\hMsun\ for the dark matter particles and
$m_{\rm gas}=1.5\times 10^{9}$\hMsun\ for the gas particles,
respectively. For more details we refer the reader to the paper that
describes the simulation and presents results drawn from it
\citep{Gottloeber07}.

For the comparison presented here we discarded the gas particles as
not all halo finders yet incorporate proper treatment of gas physics
in their codes. The focus here lies with the dark matter
structures. However, to avoid that too many particles will be
considered ``unbound'' (for those halo finders that perform an
unbinding procedure), the masses of the dark matter particles have
been corrected for this, i.e. $m_{\rm DM}^{\rm corrected}=m_{\rm
  DM}/(1-f_b)$ where $f_b=\Omega_b/\Omega_m$ is the cosmic baryon
fraction of our model universe.

In order to allow non-parallel halo finders to participate in this
test we degraded the resolution from the original $1024^3$ particles
down to $512^3$ as well as to $256^3$ particles. The properties to be
compared will however be drawn from the highest-resolved data set for
each individual halo finder, making the appropriate mass/number cuts
when producing the respective plots.

\subsection{Code Participation} \label{sec:codeparticipation}

\begin{table*}
  \caption{Brief summary of the codes participating in the comparison
    project. The first six columns provide a synopsis of the
    respective tests the code participated in (columns 2--7). The last two columns simply list whether the code performs an unbinding procedure and provided subhalo properties, respectively.}
\label{tab:codes}
\begin{center}
\begin{tabular}{lllllllll}
\hline
code             		& \multicolumn{6}{c}{participation in test}													& unbinding	& subhaloes 	\\
				& recovery	& rad. depend. 	& dyn. infall 	& resolution 	& blind 		& cosmology			&			&			\\
\hline
 \texttt{AHF}                & yes		& yes		& yes		& yes		& yes		& 1024$^3$			& yes		& yes		\\
 \texttt{ASOHF} 		& yes		& yes		& yes		& yes		& yes		& 256$^3$			& yes		& yes		\\
 \texttt{BDM} 		& yes		& yes		& yes		& yes		& yes		& 512$^3$			& yes		& yes		\\
 \texttt{pSO} 		& only host	& no			& no			& no			& only host	& 1024$^3$			& no			& no			\\
 \texttt{LANL} 		& only host	& no			& no			& no			& no			& 1024$^3$			& no			& no			\\
 \texttt{SUBFIND} 	& yes		& yes		& yes		& yes		& yes		& 1024$^3$			& yes		& yes		\\
 \texttt{FOF} 		& yes		& yes		& yes			& yes			& no			& 1024$^3$, no \vmax	& no			& limited		\\
 \texttt{pFOF} 		& only host	& no			& no			& no			& no			& 512$^3$			& no			& no			\\
 \texttt{Ntropy-fofsv} 	& only host	& no			& no			& no			& no			& 1024$^3$, no \vmax	& no			& no			\\
 \texttt{VOBOZ} 		& yes		& yes		& no		& yes		& yes		& 512$^3$			& yes		& yes		\\
 \texttt{ORIGAMI} 	& no			& no			& no			& no			& no			& 512$^3$			& yes		& no			\\
 \texttt{SKID} 		& yes		& yes		& yes			& yes		& yes		& 1024$^3$			& yes		& yes		\\
 \texttt{AdaptaHOP} 	& yes		& yes		& yes		& yes		& yes		& 512$^3$			& no			& yes		\\
 \texttt{HOT} 		& yes		& yes		& yes		& yes		& yes		& no					& no			& yes		\\
 \texttt{HSF} 		& yes		& yes		& yes		& yes		& yes		& 1024$^3$			& yes		& yes		\\
 \texttt{6DFOF}		& yes		& yes		& yes		& yes		& yes		& 1024$^3$			& no			& yes		\\
 \texttt{Rockstar}    & yes & yes & yes & yes & no & 1024$^3$ & yes & yes \\
\hline
\end{tabular}
\end{center}
\end{table*}

Not all codes have participated in all the tests just introduced and
outlined. Hence in order to facilitate an easier comparison of the
results and their relation to the particular code we provide in
\Tab{tab:codes} an overview of the tests and the halo finders
participating in them. In that regard we also list for the
cosmological simulation the respective resolution of the data set
analysed by each code. The last two columns simply indicate whether the
code performs an unbinding procedure and provided subhalo properties,
respectively.

\section{The Comparison} \label{sec:comparison}
This Section forms the major part of the paper as it compares the halo
catalogues derived with various halo finders when applied to the suite
of test scenarios introduced in the previous Section. We first address
the issue of the controlled experiments brought forward in
\Sec{sec:mockhaloes} followed by the analysis of the cosmological
simulation introduced in \Sec{sec:cosmologicalsimulation}. As already
mentioned before, we are solely addressing dark matter haloes leaving
the inclusion of baryonic matter (especially gas) for a later study.

\subsection{Mock Haloes} \label{sec:mockhaloes}
Before presenting the results of the cross comparison we need to
explain further the actual procedures applied. Each data set was given
to the respective code representative asking them to return the centre
of each object found as well as a list of the (possible) particles
belonging to each (sub)halo. A single code only using that particular
list was then used to derive the bulk velocity $V_{\rm bulk}$, the
(fiducial) mass $M_{200}$, and the peak of the rotation curve \vmax\
in order to eliminate differences in the determination of said values
from code to code. Or in other words, we did not aim at comparing how
different codes calculate, for instance, \vmax\ or $M_{200}$ and so
eliminated that issue. This simple analysis routine is also available
from the project website. We were aiming at answering the more
fundamental question ``Which particles may or may not belong to a
halo?'' according to each code. However not all representatives
returned particle lists as requested (due to a different method or
technical difficulties) but rather directly provided the values in
question; those codes are \texttt{BDM}, \texttt{FOF}, and
\texttt{6DFOF}. Further, \texttt{FOF} did not provide values for
\vmax.

And when comparing results we primarily focused on fractional
differences to the theoretical values by calculating $\Delta x/x_{\rm
  Model} = (x_{\rm code}-x_{\rm Model})/x_{\rm Model}$ where $x$ is the
halo property in question.

\subsubsection{Recovery of Host and Subhalo Properties} \label{sec:static}
For all the subsequent analysis and the plots presented in this
subsection~\ref{sec:static} we used the the setups (i) through (iii)
specified in \Sec{sec:mockdata}. In that regard we have three host
haloes (one for the host alone, one from the host+subhalo setup, and
one from the host+subhalo+subsubhalo configuration); we further have
two subhaloes at our disposal (one from the host+subhalo and one from
the host+subhalo+subsubhalo tests) as well as one subsubhalo. In all
figures presented below the origin of the halo is indicated by the
size of the symbol: the largest symbol refers to the
host+subhalo+subsubhalo set with the symbol size decreasing in the
order of the host+subhalo towards the host test alone. We further
always show the results for the NFW mock haloes in the left panel and
the Plummer spheres in the right one. As much as possible, the halo
finders have been organised in terms of their methodology: spherical
overdensity finders first followed by FOF-based finders with 6D
phase-space finders last.

\begin{figure}
  \psfig{figure=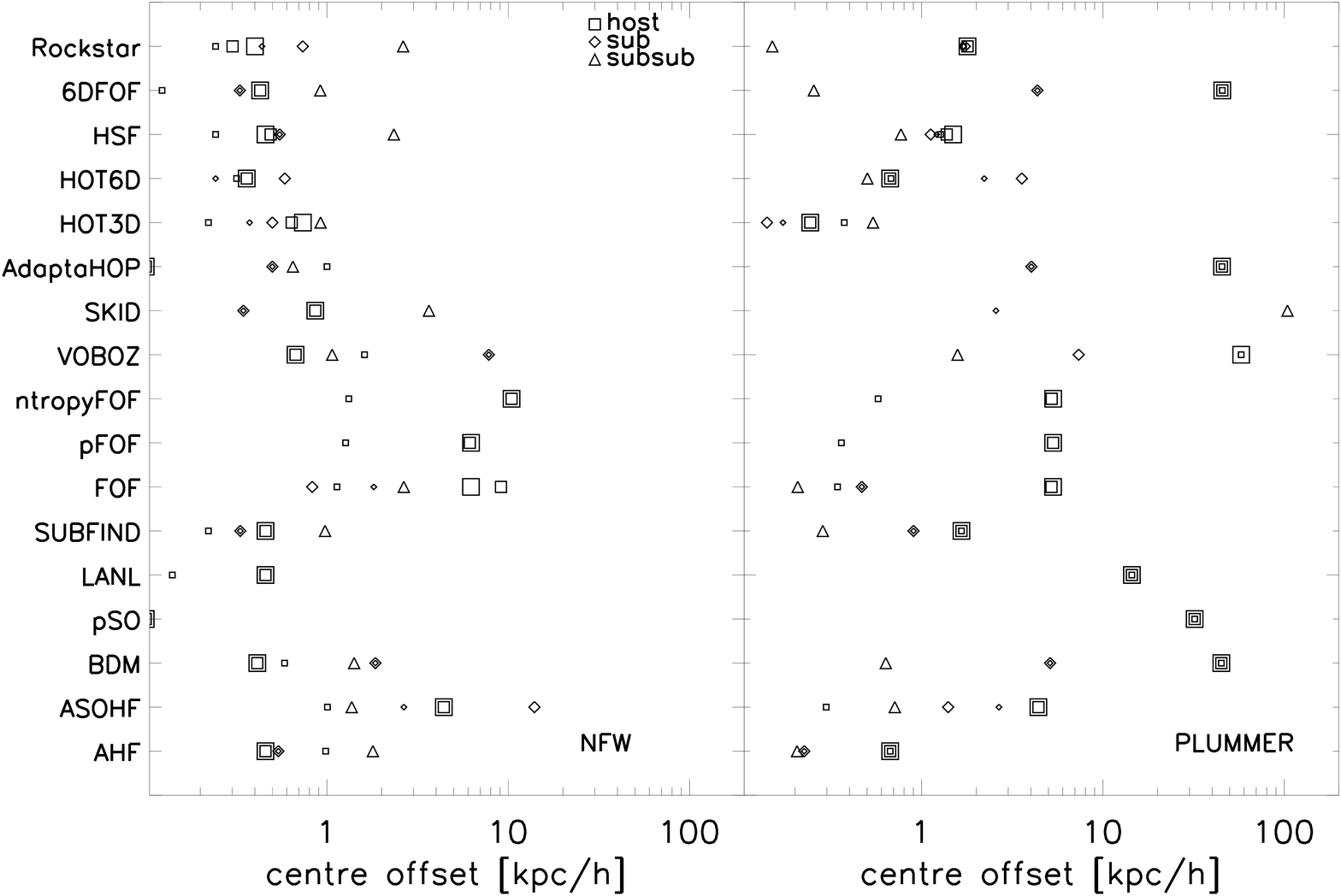,width=\hsize,angle=0}
  \caption{The offset of the actual and recovered centres for the NFW
    (left) and Plummer (right) density mock haloes. The symbols refer
    to either the host halo, subhalo or subsubhalo as indicated while
    the symbol size indicates the test sequence as detailed in the
    text (i.e. larger symbols for haloes containing more subhaloes).}
\label{fig:MOCK-centreoffset}
\end{figure}

\paragraph*{Centre Determination}
We start with inspecting the recovery of the position of the haloes as
practically all subsequent analysis as well as the properties of
haloes depend on the right centre determination. The results can be
viewed in \Fig{fig:MOCK-centreoffset} where the $y$-axis represents
the halo finder and the $x$-axis measures the offset between the
actual position and the recovered centre in \hkpc.
 
We can clearly see differences for all sorts of comparisons: host
haloes vs. (sub-)subhaloes, NFW vs. Plummer model, and -- of course --
amongst halo finders. While for the NFW density profile the deviations
between analytical and recovered centre are for the majority of haloes
and codes below $\approx$5\hkpc\ there are nevertheless some
outliers. For the large halo the $100^{th}$ particle is $3$\hkpc\ from
the nominal centre. These outliers are primarily for the FOF-based
halo finders which are using a centre-of-mass rather than a
density-peak as the centre. However, for a perfectly spherically
symmetric setup as the one used here the differences between
centre-of-mass and density peak should be small. Some of the finders
(\texttt{pSO, LANL, pFOF, ntropy-fofsv}) were not designed to find
substructure and so do not return the locations for
these. Interestingly \texttt{HOT6D} cannot detect the NFW
subsubhalo. The situation is a bit different for the Plummer model
that consists of a flat density profile inwards from the scale radius
of 190\hkpc. While the centre offset for the FOF finders remains the
same we now also observe a shift towards larger offset-values for the
majority of the other codes; some codes were even unable to locate the
host halo at all (e.g. \texttt{SKID}) while other finders even
marginally improved their (sub-)halo centre determination
(\texttt{AHF}, \texttt{ASOHF}, \texttt{HOT3D}). Remember that for
\texttt{6DFOF} all positions and velocities were solely determined
from the linked particles which explains the slightly offset centres
in cases only a few particles were linked (as in the case of the
Plummer sphere which had an artificial low phase space density by
construction) as well as the somewhat different bulk velocities (when
compared to the other halo finders below).

\begin{figure}
  \psfig{figure=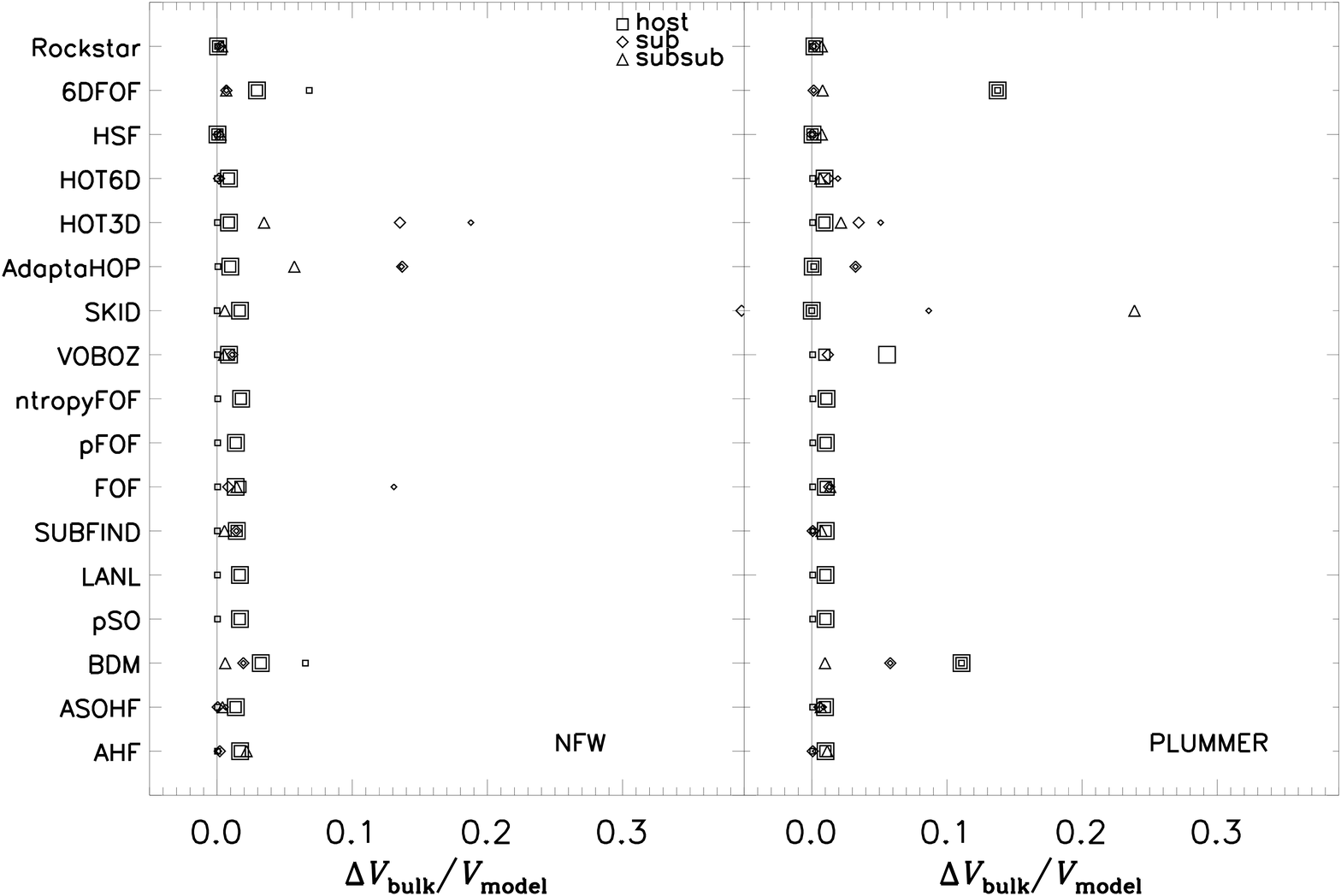,width=\hsize,angle=0}
 \caption{Recovery of halo bulk velocities in comparison to the
    analytical input values for the NFW (left) and Plummer (right)
    density mock haloes. Note that the host halo has been set up to be
    at rest with $v_{\rm bulk}=0$. The symbols have the same meaning
    as in \Fig{fig:MOCK-centreoffset}}
\label{fig:MOCK-Vbulk}
\end{figure}

\paragraph*{Halo Bulk Velocity}
A natural follow-up to the halo centre is to ask for the credibility
of the bulk velocity of the halo. Errors in this value would indicate
contamination from particles not belonging to the halo in question to
be studied in greater detail in \Sec{sec:resolution} below. In our
test data the host is always at rest whereas the subhalo (subsubhalo)
flies towards the centre with -1000 (-1200) km/sec along the negative
$x$-direction. The fractional difference between the model velocity
and the bulk velocity as measured for each halo finder is presented in
\Fig{fig:MOCK-Vbulk}. Please note that we have normalised the host's
velocities to the rotational velocity at the $R_{100}$,
i.e. $\approx$1000 km/sec, for the two density profiles. Here we find
that for practically all halo finders the error in the bulk velocity
is smaller than 3\%; only some outliers exist. Please note that we
used all particles in the determination of the bulk velocities as
returned/recovered by the respective halo finder. \texttt{SKID}
displays very significant contamination in the recovered subhaloes
with a 40\% error in the recovered bulk velocity but is also one of
the codes whose returned particle lists are intended to undergo
significant post-processing. \texttt{AdaptaHOP} and \texttt{HOT3D}
have smaller but still significant levels of contamination within the
returned substructures. The marginal offset in the bulk velocities of
the host Plummer host haloes for \texttt{6DFOF} and \texttt{BDM} is
directly related to the respective centre offsets seen in
\Fig{fig:MOCK-centreoffset}: those two codes base their bulk
velocities on particles in the central regions.

\begin{figure}
  \psfig{figure=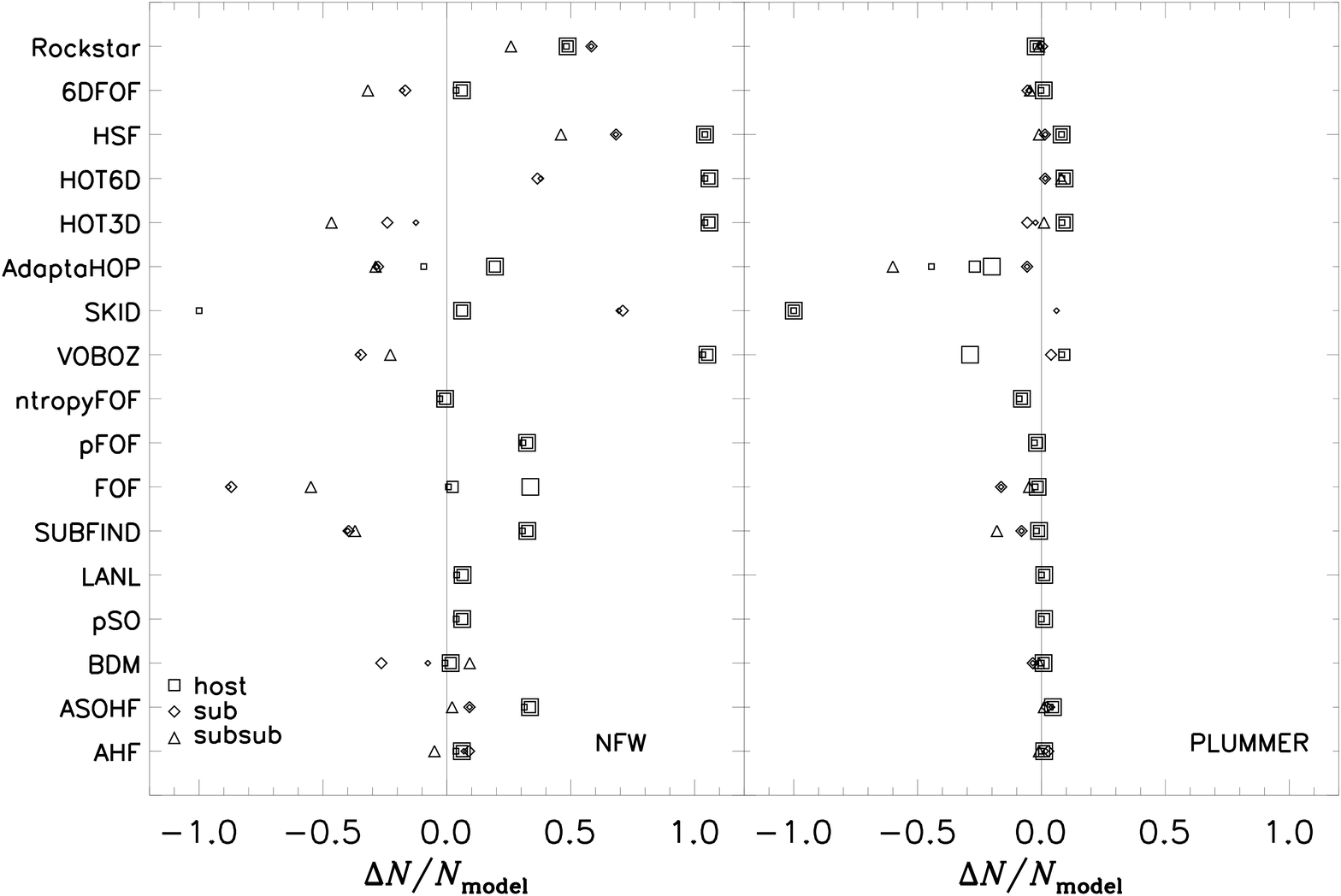,width=\hsize,angle=0}
 \caption{Total number of particles recovered for the (sub-)halo for the NFW
    (left) and Plummer (right) density mock haloes with respect to the number
    of particles within $M_{200}$. The symbols have the same meaning as in
    \Fig{fig:MOCK-centreoffset}}
\label{fig:MOCK-Npart}
\end{figure}

\paragraph*{Number of Particles}
In \Fig{fig:MOCK-Npart} we are comparing the number of particles
recovered by each halo finder to the number of particles within
$M_{200}$ listed in \Tab{tab:static}\footnote{Please note that in all
  subsequent plots we are using $N_{200}$ when referring to $N_{\rm
    model}$.}. We are aware that there is no such well defined radius
for (sub-)subhaloes, but it nevertheless provides a well-defined base
to compare against.

We observe that while the errors are at times substantial for the NFW
model the Plummer results appear to be more robust this time. But this
is readily explained by the form of the applied density profile: the
variations in mass and hence number of particles are more pronounced
for the NFW profile than for the Plummer model when changing the
(definition of the) edge of a halo. Or in other words, the total mass
of a Plummer model is well-defined whereas the mass of an NFW halo
diverges. Therefore, (minor) changes and subtleties in the definition
of the other edge of a (sub-)halo will lead to deviations from the
analytically expected value -- at least for the NFW model. To this
extent we also need to clarify that each halo finder had been asked to
return that set of particles that was believed to be part of a
gravitationally bound structure; participants were not asked to return
the list of particles that made up $M_{200}$. Post-processing of the
supplied particle lists to apply this criterion results in errors for
the NFW profiles that are well below 10\% -- at least for the host
haloes (cf. \Fig{fig:MOCK-M200} below). However, a straight comparison
of the number of recovered particles amongst the codes reveals
a huge scatter. This is due to the fact that the individual codes are
tuned to different criteria to define the edge of the halo. Clearly
some codes (\texttt{HSF, HOT, VOBOZ}) have been tuned to extract an
effectively smaller overdensity for this test than say \texttt{6DFOF,
LANL, pSO} or \texttt{AHF}. This is a well known issue and all code
developers are well aware of it. Perhaps more concerning is the wide
scatter in relative mass of the largest subhalo. Here $M_{200}$ is
ill-determined but the ratio of the substructure mass to the host halo
mass displays a wide scatter. This ratio is a astrophysical importance
for several issues.

The difference in host halo seen for \texttt{FOF} and \texttt{pFOF} is
-- in general -- due to the choice of a linking-length not
corresponding to $200\times \rho_{\rm crit}$. However, with an
appropriate linking length the \texttt{FOF} algorithm detects the halo
at the desired overdensity correctly as can be seen for the host only
and host+subhalo data for which there is agreement with the analytical
expectation as opposed to the host+subhalo+subsubhalo where the
standard linking length has been applied and hence the number of
particles (and mass) is over-estimated. As a (down-)tuned linking
length has also been utilized for the detection of the (positions of
the) subhaloes, the higher overdensity encompassed naturally led to a
smaller number of particles (and masses) than assumed in the model.

Again, we stress that \Fig{fig:MOCK-Npart} does not necessarily
reflect the number of particles actually used to calculate halo
properties; it is the raw number of (bound) particles assigned to the
centre of the respective (sub-)halo and used for further
post-processing with most of the codes. But the comparison also
indicates that neither number of particles nor $M$ as defined by some
overdensity criterion (see below) are stable quantities for a fair
comparison; this is why we argue in favour of the peak of the rotation
curve for cross-comparison as already highlighted in the
introduction.

\paragraph*{Mass}
Using the particle lists provided by each halo finder we extract each
object and calculate the density profile. From this we determine the
point where it drops below $200\times\rho_{\rm crit}$. This point can
then be used as a radial distance within which to define $M_{200}$
which is then compared against the theoretical expectation
(cf. \Tab{tab:static}) in \Fig{fig:MOCK-M200}. Again, we acknowledge
that this is not the correct definition for (sub-)subhalo mass, but
can regardlessly be used to compare halo finders amongst themselves.

As already outlined in the previous paragraph, the differences to the
analytical values (and between the codes) are substantially alleviated
now that differences in definition for the edge of each halo have been
removed. The apparent underestimation of the (sub-)subhalo masses has
also to be taken and digested carefully as the $M_{200}$ values are based
upon objects in isolation when these are embedded in a large host
halo. However, please recall that the values for \texttt{BDM, FOF,
  6DFOF} are based upon their respective criteria as these codes did
not return particle lists but directly $M_{200}$.

Amongst those codes that did recover subhaloes and underwent the same
processing scheme there remains a surprisingly wide variation in
recovered subhalo $M_{200}$ mass. Almost all the codes studied here
post-process their subhalo catalogues heavily to alleviate this
problem. We would stress however that the precise definition for a
subhalo contents can, as demonstrated, lead to a range of recovered
subhalo masses, a point users of subhalo catalogues should be well
aware of. We will return to the issue of missing subhalo mass in
\Sec{sec:dynamicinfall} below, which provides some explanation for the
variation.

\begin{figure}
  \psfig{figure=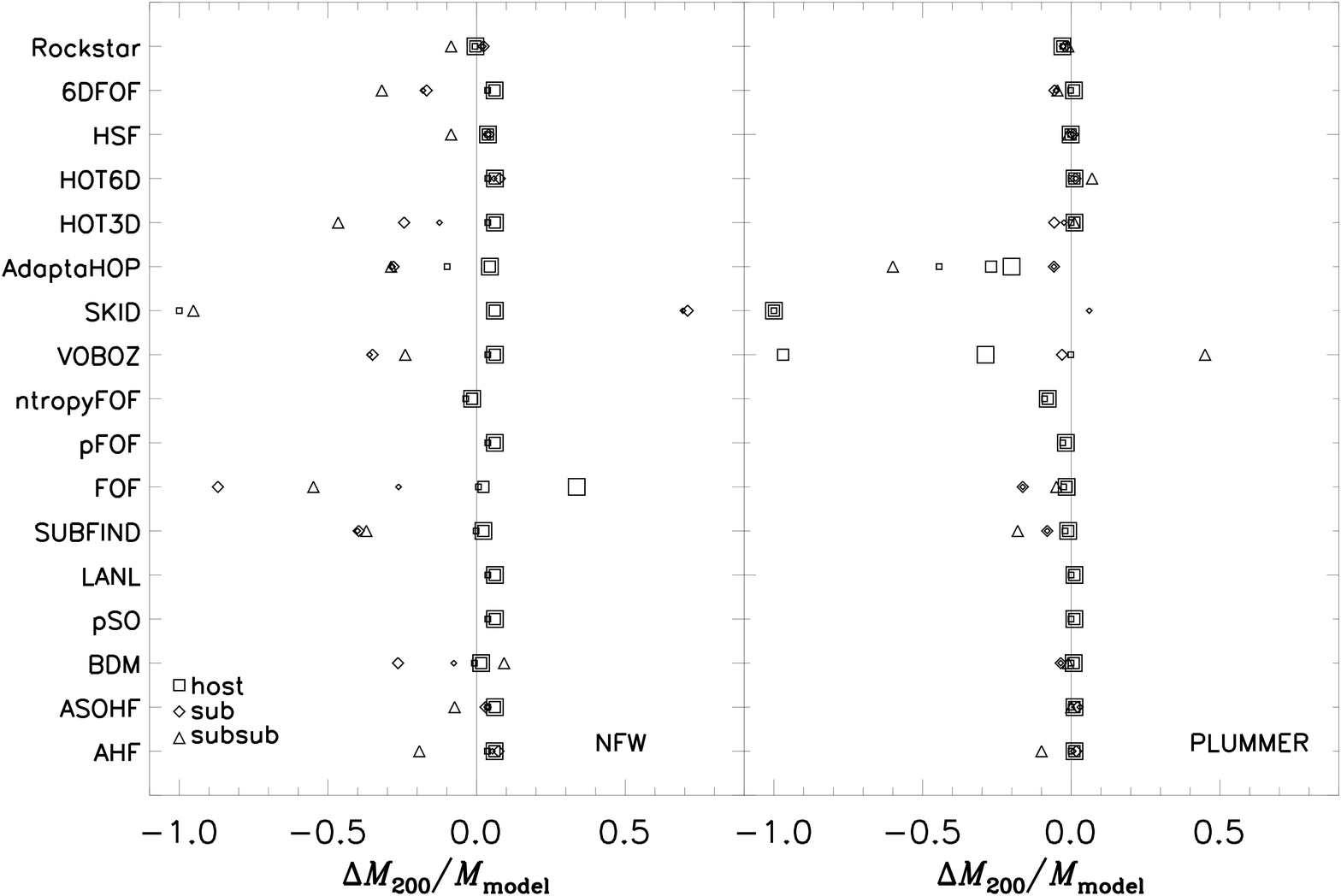,width=\hsize,angle=0}
  \caption{$M_{200}$ mass (as determined from the supplied particle
    lists) measured according to the mean enclosed density being
    $200\times\rho_{\rm crit}$ criterion for the NFW (left) and
    Plummer (right) density mock haloes extracted from each finder's
    list of gravitationally bound particles. The symbols have the same
    meaning as in \Fig{fig:MOCK-centreoffset}}
\label{fig:MOCK-M200}
\end{figure}

\begin{figure}
  \psfig{figure=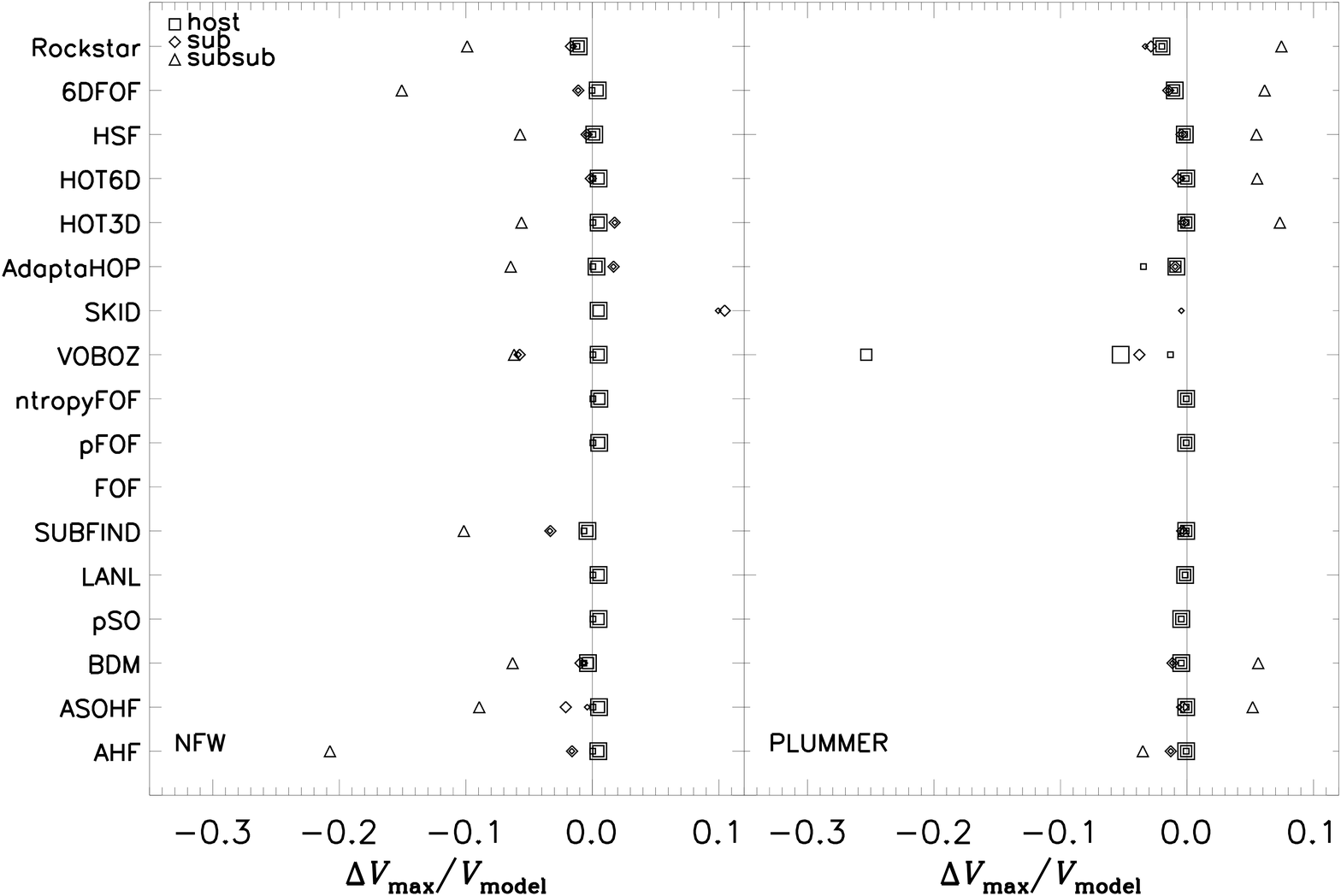,width=\hsize,angle=0}
 \caption{Recovery of numerical \vmax\ values in comparison to the
    analytical input values for the NFW (left) and Plummer (right)
    density mock haloes.The symbols have the same meaning
    as in \Fig{fig:MOCK-centreoffset}}
\label{fig:MOCK-Vmax}
\end{figure}

\paragraph*{Maximum of the Rotation Curve}
As outlined in \Sec{sec:howtocomparehaloes}, $M_{200}$ does not
provide a fair measure for (sub-)subhalo mass and hence we consider
the maximum circular velocity \vmax\ as a proxy for mass. The
fractional difference between the theoretically derived \vmax\ and the
value based upon the particles returned by each halo finder are
plotted in \Fig{fig:MOCK-Vmax}. While we now find a considerably
improved agreement with the analytical calculation the subsubhalo has
still not been recovered correctly in most of the cases. This result
is entirely in line with the results of figure 7 of \citet{Muldrew11}
where the error in measuring \vmax\ for a range of particle numbers
was calculated: we should not be surprised by a 10\%
underestimate for our subsubhalo as this is well within expected
limits.

\subsubsection{Radial Dependence of Subhalo Properties} \label{sec:radialdependence}
\begin{figure}
  \psfig{figure=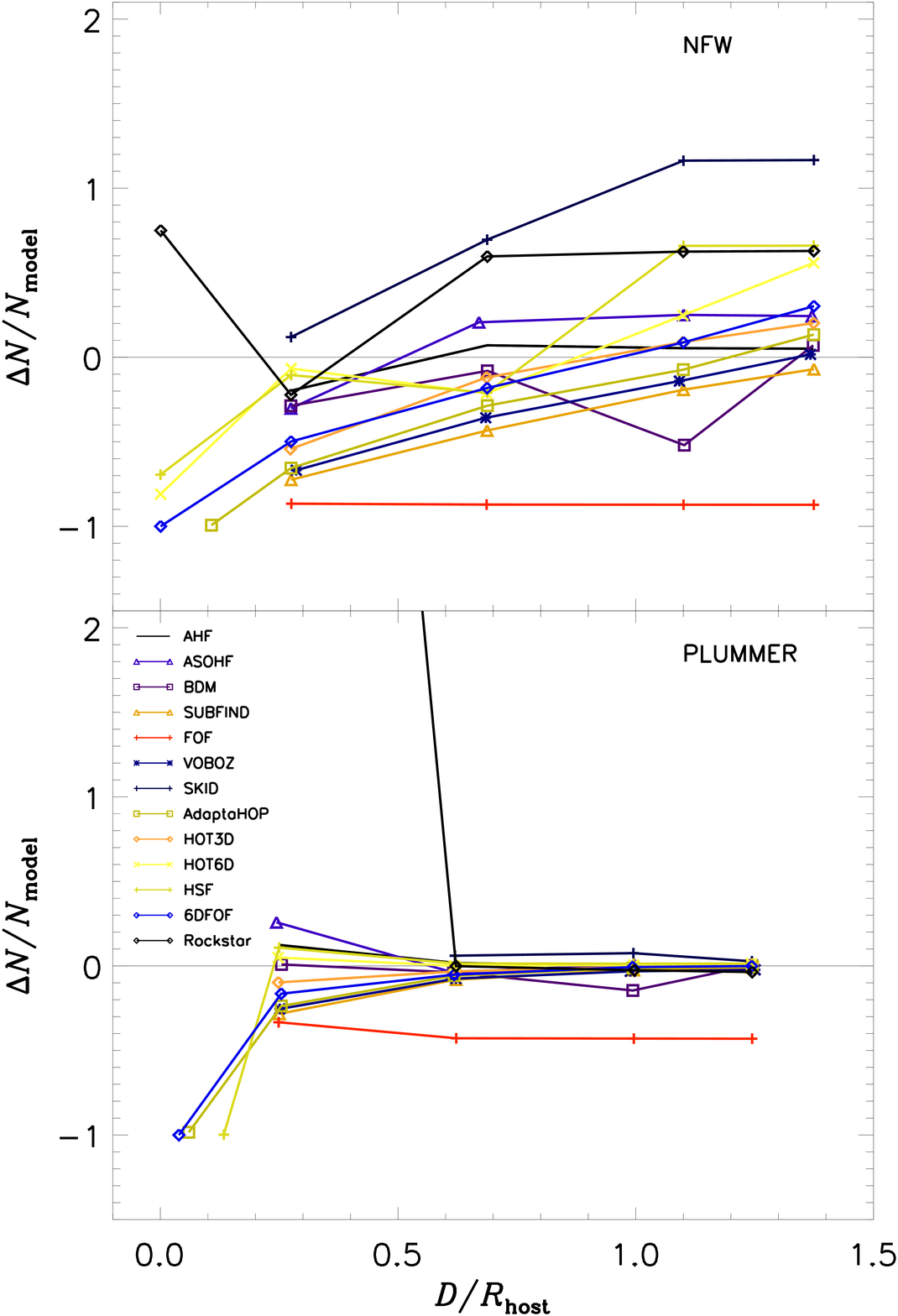,width=\hsize,angle=0}
  \caption{Number of particles belonging to the subhalo for the NFW
    (upper) and Plummer (lower) density mock haloes as a function of
    subhalo distance to the host.}
\label{fig:MOCK-NpartDist}
\end{figure}

\begin{figure}
  \psfig{figure=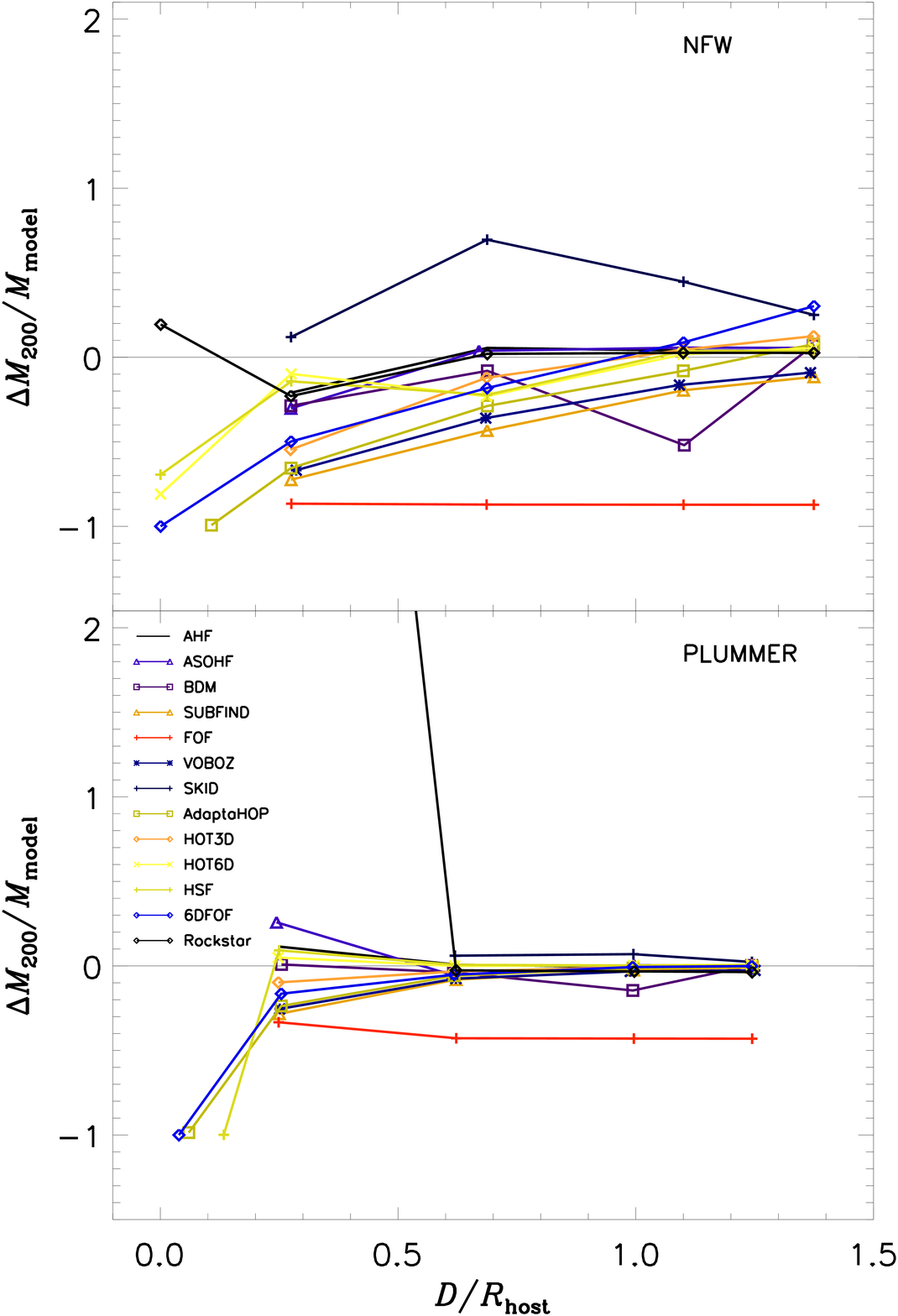,width=\hsize,angle=0}
  \caption{Hypothetical $M_{200}$ value comparison to the NFW (upper)
    and Plummer (lower) subhalo as a function of distance to the
    host. $M_{200}$ was calculated again considering the recovered
    particles $N$ (as presented in \Fig{fig:MOCK-NpartDist}) in
    isolation. }
\label{fig:MOCK-M200Dist}
\end{figure}

\begin{figure}
  \psfig{figure=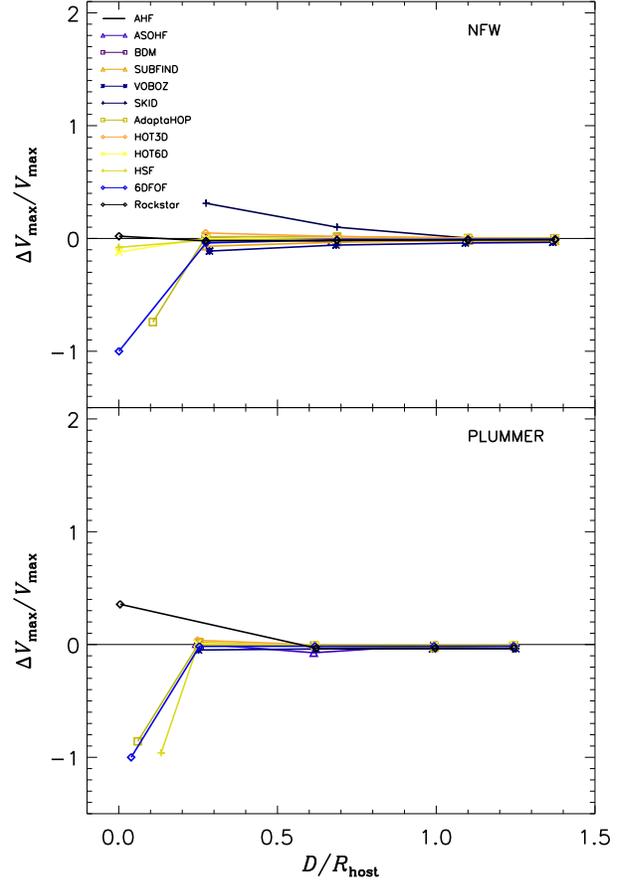,width=\hsize,angle=0}
 \caption{Recovery of numerical \vmax\ values in comparison to the
    analytical input values for the NFW (upper) and Plummer (lower)
    density mock haloes as a function of subhalo distance to the
    host.}
\label{fig:MOCK-VmaxDist}
\end{figure}

The following test aims at studying how the recovered properties of a
subhalo change as a function of the distance from the centre of the
subhalo to the centre of its host. We always placed the same subhalo
(sampled with 10000 particles) at various distances and applied each
halo finder to this test scenario, without changing the respective
code parameters in-between the analyses. We then focused our
attention on the number of gravitationally bound particles in
\Fig{fig:MOCK-NpartDist}, the recovered $M_{200}$ masses in
\Fig{fig:MOCK-M200Dist} and the maximum of the rotation curve in
\Fig{fig:MOCK-VmaxDist}.

We reiterate that this particular test (as well as the following two)
is only suited to halo finders that are able to identify substructure
embedded within the density profile of a larger encompassing
object. Therefore, some of the codes will not appear in this and the
following tests in \Sec{sec:resolution} and
\Sec{sec:dynamicinfall}. However, we also need to acknowledge that
some of the code developers were keen to participate in this venture
and manually tuned their halo finders to (at least) provide a centre
(and possibly mass) estimate for the subhalo under investigation
(e.g. \texttt{FOF} by Gottl\"ober \& Turachninov systematically
lowered their linking length until an object had been found using the
spin parameter as a measure for credibility (cf. \Sec{sec:fof});
however, as \texttt{FOF} in its basic implementation does not perform
any unbinding they did not dispense particle lists and/or internal
properties.). Therefore, the results for \texttt{FOF} are to be taken
lightly and with care.

\paragraph*{Number of Particles}
Aside from the location of the substructure, which we are not investigating in more
detail in this particular Subsection, the number of particles
recovered by each halo finder is the first quantity to explore as a
function of subhalo distance. The results can be viewed in
\Fig{fig:MOCK-NpartDist} with the NFW mock halo in the upper panel and
the Plummer sphere in the lower. Recall that there are five subhaloes
placed at various distances from the centre of the host with the
closest one actually overlapping with the host centre. 

As expected from the above results of the previous section (which
equate to the middle position of these five haloes) the various halo
finders recover a range of number of particles within the halo. Only
the phase space based finders are capable of disentangling the subhalo
when it is directly in the centre. Even then their particle recovery
either indicates that there are too few particles associated with the
subhalo or that they found the host. We further observe that, at least
for the NFW haloes, the number of recovered particles drops the closer
we get to the centre. This is naturally explained by the fact that the
density contrast of the subhalo becomes smaller and the point where
the host halo's density takes over is closer to the centre of the
subhalo. This is another reflection of the fact that the number of
particles (or anything based upon a measure of ``halo edge'') is not a
good proxy for the actual subhalo. The situation is obviously
different for the Plummer sphere with no pronounced density rise
towards the centre; therefore, the subhalo appears to be well
recovered in this case. For the low number of particles recovered by
\texttt{SUBFIND} we refer the reader to an improved discussion and
investigation, respectively, in \citet{Muldrew11}.

In any case, these are the still simply particle lists; we continue to
check the (hypothetical) $M_{200}$ values as well as the recovery of
the maximum of the rotation curve.  When defining a (hypothetical)
$M_{200}$ value considering the subhalo in isolation we find basically
the same trends as for the number of particles. This can be verified
in \Fig{fig:MOCK-M200Dist} where we observe the same phenomena as in
\Fig{fig:MOCK-NpartDist}. However, \texttt{SKID} is the exception with
the $M_{200}$ values closer to the actual model mass across all
distances than the number of particles, as expected and as they
themselves would obtain during their own post-processing steps.

We note that the discrepancy between the (fiducial) mass and the real
mass of the subhalo placed at different radial distances from the
centre is more serious in this idealised set-up than it would be in a
realistic situation, where the substructures would experience tidal
truncation in moving towards the inner regions of the halo (see the
discussion in \Sec{sec:mockdata} as well as the study of the dynamical
subhalo infall in \Sec{sec:dynamicinfall} below); when considering the
mass within the tidal truncation radius, the discrepancy between the
``real'' and recovered mass would reduce.

The most credible measure of subhalo mass, however, appears to be the
maximum of the rotation curve: it hardly changes its value
irrespective of the position inside the host halo as can be seen in
\Fig{fig:MOCK-VmaxDist}. All halo finders perform equally well in
recovering the \vmax\ value from the list of particles used in
\Fig{fig:MOCK-NpartDist}. This then indicates that the only difference
amongst the halo finders as seen as a substantial spread in (the upper
panel of) \Fig{fig:MOCK-NpartDist} stems from the outer and less well
contrasted regions of the subhalo.

\paragraph*{Maximum of Rotation Curve}
We have seen in \Sec{sec:static} that the maximum of the rotation
curve \vmax\ serves as an adequate proxy for mass and hence we test
its sensitivity to radial position in \Fig{fig:MOCK-VmaxDist}. We find
that this quantity is, as expected, hardly affected by the actual
position of the subhalo within the host. Its value is determined by
the more central regions of the subhalo and hence does not change if
the object is truncated in the outskirts due to embedding within the
host's background density field. Only when the two centres of the sub-
and the host halo overlap do we encounter problems again, however,
\texttt{HOT6D} and \texttt{HSF} even masters this situation fairly
well (at least for the more realistic NFW test scenario).

\subsubsection{Dynamical Infall of a Subhalo} \label{sec:dynamicinfall}
The test described and analysed in this Subsection is a dynamic
extension of the previously studied radial distance test: we throw a
subhalo (initially sampled with 10000 particles inside $M_{100}$) into
a host halo two orders of magnitude more massive. It was initially
placed at a distance of $D=3\times R_{100}^{\rm host}$ with a radially
inwards velocity of $v=\sqrt{2GM(<D)/D}=686$km/s and then left to
free-fall. During the temporal integration of this system with
GADGET-2 the cosmological expansion was turned off so the haloes were
only affected by gravity. The orbit of the subhalo takes it
right through the host halo centre, exiting on the other side. Due to
the tidal forces the subhalo will lose mass and we aim at quantifying
how different halo finders recover both the number of (bound)
particles as well as the evolution of the peak rotational velocity.

\begin{figure}
  \psfig{figure=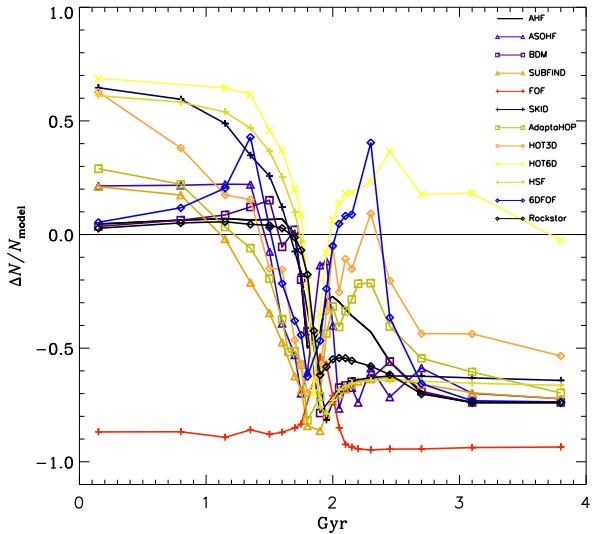,width=\hsize,angle=0}
  \caption{Temporal evolution of the number of particles belonging to
    the subhalo for the dynamical infall study.}
\label{fig:DYN-Npart}
\end{figure}

\paragraph*{Evolution in Number of Particles}
In \Fig{fig:DYN-Npart} we start again with the number of recovered
particles this time as a function of time measured in Gyrs since the
infalling object passed $2\times R_{200}^{\rm host}$. Note the
fractional difference $\Delta N/N_{\rm model}$ is measured with
respect to the number of particles $N_{\rm model}$ prior to infall and
that the analysis has only been performed over a certain number of
output snapshots and not every integration step. At the starting point
we observe again the same scatter in the number of particles as
already found in \Fig{fig:MOCK-NpartDist}.\footnote{However, when
  comparing \Fig{fig:MOCK-NpartDist} and \Fig{fig:DYN-Npart} one needs
  to bear in mind that the radial dependence of subhalo properties
  only extends out to $\approx 1.37\times R_{200}^{\rm host}$ whereas
  the first data point in \Fig{fig:DYN-Npart} is for $2\times
  R_{200}^{\rm host}$.} Until the passage through the very centre of
the host halo after approximately 1.8 Gyrs we also find the expected
drop in number of particles due to the stripping of the subhalo;
however, as noted in \Fig{fig:MOCK-NpartDist} part of this drop can
also be attributed to the subhalo moving deeper into the dense region
of the host. This drop in particle number has a marginally different
shape depending on the halo finder, and for \texttt{ASOHF} there is
even a marginal rise. But this time actually all halo finders (except
the phase-space finders \texttt{HOT6D}, \texttt{HSF} and
\texttt{6DFOF}, cf. \Fig{fig:DYN-VmaxZoom} below) do lose the subhalo
when it overlaps with the host halo - or at least are unable to
determine its properties at that time (e.g. \texttt{6DFOF} actually
found the objects but could not assign the correct particles to it as
the search radius for ``subhalo membership'' was practically
zero). After the passage through the centre all halo finders identify
the object again with more particles yet obviously not reaching the
original level anymore.

However, we also like to mention that after the core transition of the
subhalo we expect to find a more or less constant set of particles
that remain bound to the subhalo: as the radial distance increases
again there is no reason for the subhalo to lose additional mass. It
seems clear that the majority of structure finders agree on this
plateau value, but there are also some that return an unphysical
result in this regime (e.g. both \texttt{HOT} codes as well as
\texttt{6DFOF} in the early phases).

Please note again that none of the FOF-based halo finders is ab initio
designed to locate substructure, but the \texttt{FOF} results have
been included as this code was manually tuned to locate subhaloes
(cf. \Sec{sec:radialdependence}).

\begin{figure}
  \psfig{figure=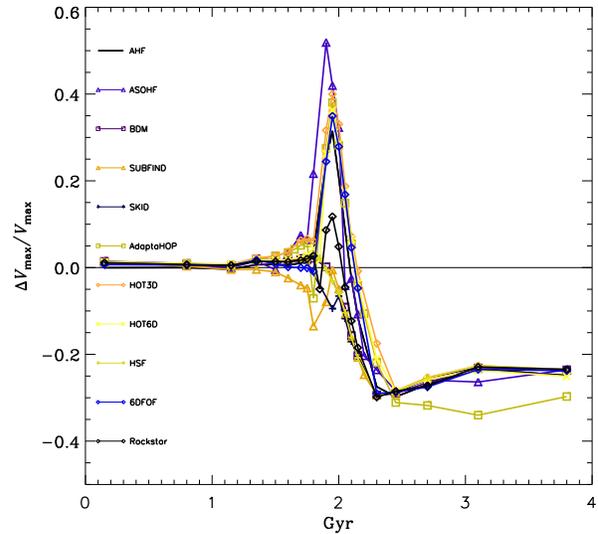,width=\hsize,angle=0}
  \caption{Temporal evolution of the maximum of the rotation curve for
    the dynamical infall study.}
\label{fig:DYN-Vmax}
\end{figure}

\paragraph*{Evolution of the Maximum of the Rotation Curve}
As we have seen before a number of times already, the number of
particles has to be used with care as the actual halo properties will
be based upon them, but the list has undeniably to be pruned and/or
postprocessed. We therefore present in \Fig{fig:DYN-Vmax} again the
evolution of the maximum of the rotation curve which focuses on the
more central regions of the subhalo and its particles. Here we can
undoubtedly see that all halo finders perform equally well (again):
they all start with a value equal to the analytical input value and
drop by the same amount once the subhalo has left the very central
regions again. However, the majority of the codes (except
\texttt{SUBFIND}, \texttt{HSF}, and \texttt{SKID}) found a sharp rise of \vmax\ right
after the central passage. 

To gain better insight into this region we show in
\Fig{fig:DYN-VmaxZoom} a zoom into the timeframe immediately
surrounding the central passage, this time though using the distance
(as measured by the respective halo finder) to the host centre as the
$x$-axis. We attribute part of this rise to an inclusion of host
particles in the subhalo's particle list to be studied in greater
detail below in \Sec{sec:resolution}; we can see that codes having
problems with such contamination appear to show this rise too -- even
though not all of the codes showing this rise are amongst the list of
finders showing contamination. However, this rise is also (or maybe
even more) indicative of problems with the unbinding procedure:
particles who have just left the subhalo (and are then part of the
host) may still be considered bound depending on the particulars of
the halo finder. For instance, \texttt{AHF} assumes a spherically
symmetric object during the unbinding process which is obviously not
correct for an object heavily elongated by the strong tides during the
central passage. However, one should also bear in mind that a rise in
\vmax\ also occurs when the subhalo gets (tidally) compressed and
hence $R_{\rm max}$ is lowered \citep[cf.][]{Dekel03} even though this
has not been seen in all (controlled) experiments of this kind
\citep[e.g.][]{Hayashi03, Klimentowski09}.

Finally we point out that the $x$-axis is based upon the
distance to the host centre as measured by each individual halo
finder; and it is rather obvious that all halo finders have recovered
(more or less) the same distance for the subhalo.

\begin{figure}
  \psfig{figure=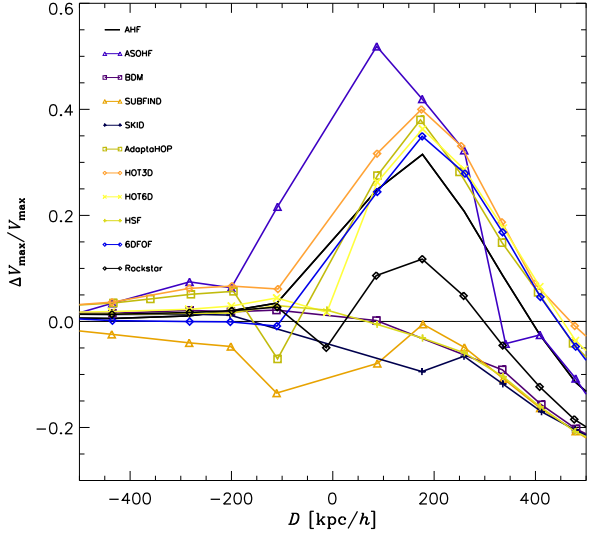,width=\hsize,angle=0}
  \caption{The maximum of the rotation curve for the dynamical infall
    study as a function of distance (as measured by the halo finder)
    to the centre of the host -- zooming into the region about the
    centre.}
\label{fig:DYN-VmaxZoom}
\end{figure}

\subsubsection{Resolution Study of a Subhalo} \label{sec:resolution}
While we have seen that there is little variation of the most stable
subhalo properties with respect to distance to the host (i.e. \vmax)
we now investigate the number of particles required to (credibly)
identify a subhalo. To this extent we used setup (ii) from the list in
\Sec{sec:mockdata} where we placed a single subhalo into a host halo
at half the host's $M_{100}$ radius. But this time we also gradually
lowered its mass and number of particles (keeping the mass of an
individual particle constant). Even though it is meaningless to talk
about $R_{200}$ radii for subhaloes again, we are nevertheless
comparing the number of gravitationally bound particles, as returned
by the respective halo finder, to the number of particles inside the
subhaloes' $R_{200}$ radius; remember that the subhaloes were
generated in isolation and sampled out to $2\times$ their $M_{100}$
radius (cf. \Sec{sec:mockdata}).

\begin{figure}
  \psfig{figure=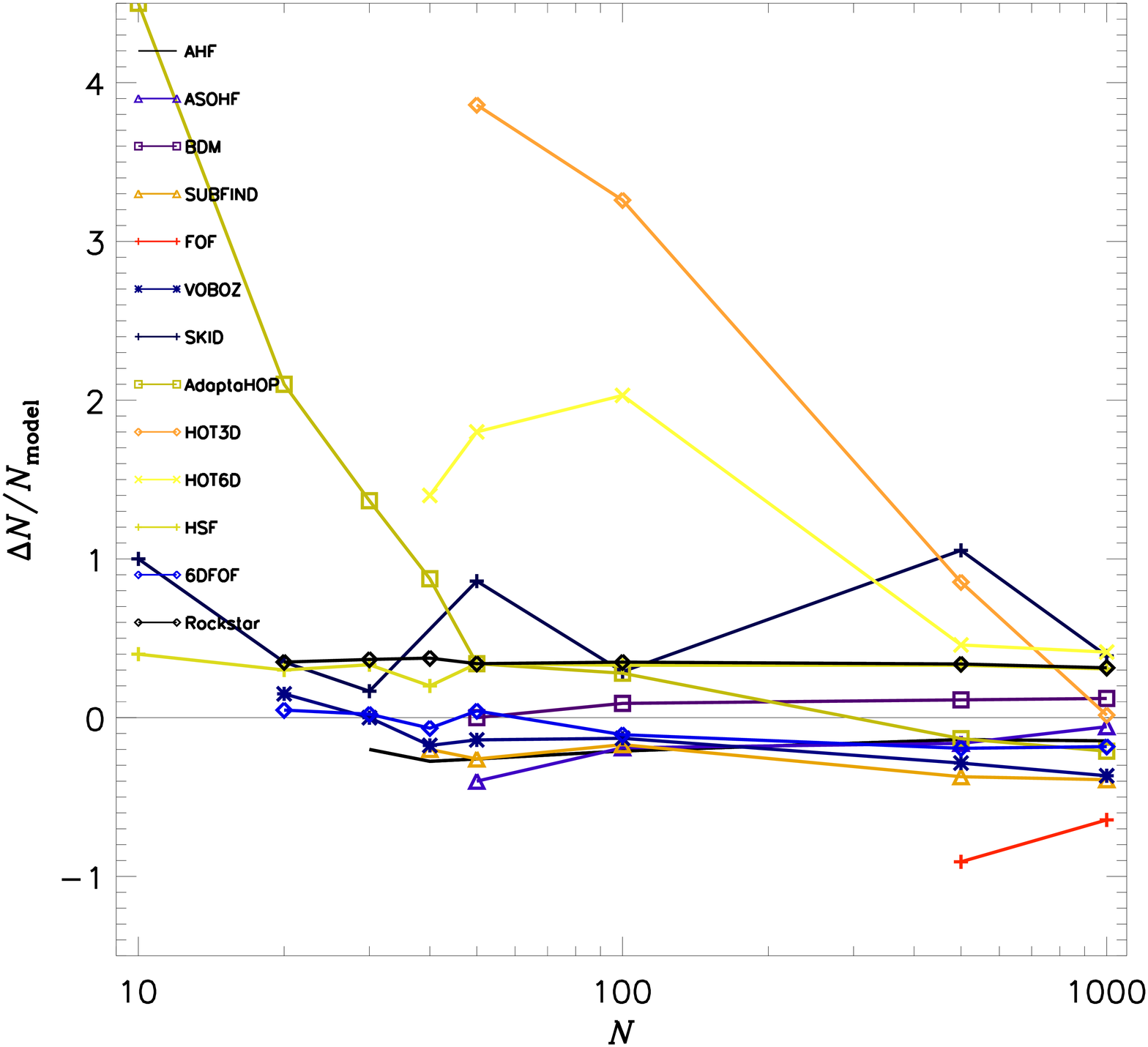,width=\hsize,angle=0}
  \caption{Fractional difference between number of particles within
    the recovered $R_{200}$ and number of particles belonging to the
    halo as returned by the respective halo finder vs. the number of
    particles inside the subhalo.}
\label{fig:RES-Npart}
\end{figure}

\paragraph*{Number of Particles}
The results of this resolution study can be viewed in
\Fig{fig:RES-Npart} where we plot the fractional difference in the
number of particles within $R_{200}$ against the number of particles
in the subhalo. In this figure there are two important things to note
and observe: a) the end point of each curve (towards lower particle
numbers) marks the point where the respective halo finder was no
longer able to identify the object and b) a constant line
(irrespective of being above, on top, or below the $0$-line) means
that for each particle number the error in the determination is
equal. Again, practically all halo finders perform equally well,
i.e. they recover the input number of particles with a constant error
across all values. Only the two \texttt{HOT} algorithms show a strong
deviation due to the lack of an unbinding procedure. It is also
interesting to compare the (inner) end point of the curves marking the
number of particles for which a certain code stopped finding the
subhalo: all of them were still able to identify the object with 50
particles. \texttt{HSF} and \texttt{SKID} actually went all the way
down to 10 particles with \texttt{VOBOZ}, \texttt{6DFOF}, and
\texttt{Rockstar} stopping at 20 particles, and \texttt{AHF} at 30. We
need to stress that codes were asked not to alter their technical
parameters while performing this resolution study and hence some may
in fact be able to recover objects with a lower number of particles
than presented here. For instance, we are aware that \texttt{SUBFIND}
(as well as \texttt{AHF} and \texttt{ASOHF}) is capable of going all
the way down to 20 particles, if the technical parameters are adjusted
appropriately.

In any case, we also observe that some codes show a rise in $\Delta
N/N_{\rm model}$ towards lower particle numbers
(e.g. \texttt{AdaptaHOP}, \texttt{HOT}); could this be due to
contamination from host halo particles? We will study this phenomenon in the
following Subsection.

\begin{figure}
  \psfig{figure=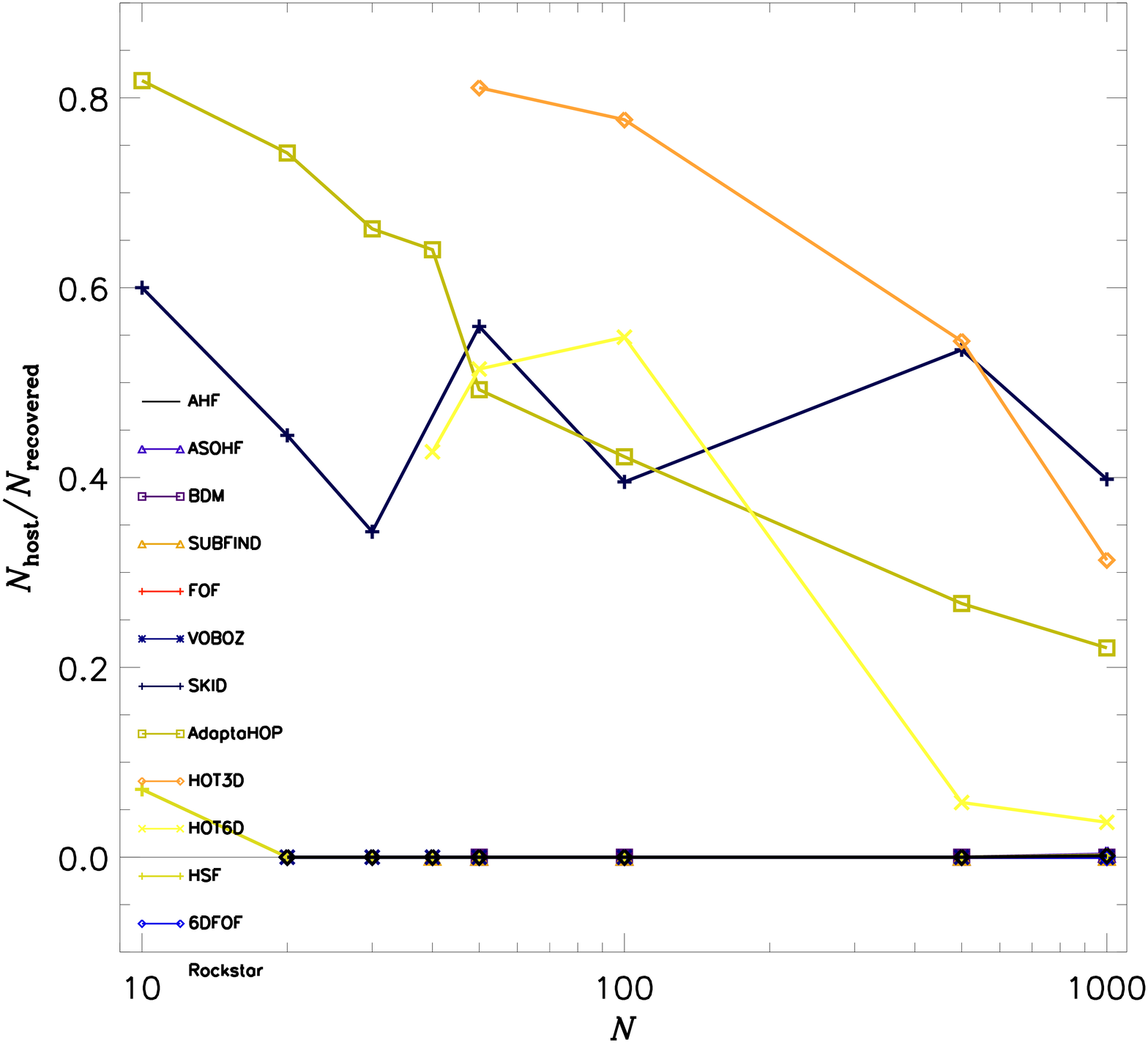,width=\hsize,angle=0}
  \caption{Fraction of host's particles identified to be part of the
    subhalo as a function of particles inside the subhalo.}
\label{fig:RES-Nhost}
\end{figure}

\paragraph*{Contamination by Host Particles}
Downsizing a subhalo yet still trying to pin-point it also raises the
question how many of the recovered particles are actually subhalo and
how many are host halo particles. We are in the unique situation to
know both the id's of the sub- and the host halo and hence studied the
``contamination'' of the subhalo with host particles as a function of
the number of (theoretical) subhalo particles in
\Fig{fig:RES-Nhost}. We can see that the vast majority of the halo
finders did not assign any host particles to the subhalo. However,
some halo finders appear to have picked up a fraction of host
particles possibly leading to differences in the subhalo properties
such as \vmax\ investigated next. Note that the high contamination for
\texttt{AdaptaHOP} is due to the lack of an unbinding procedure.

\begin{figure}
  \psfig{figure=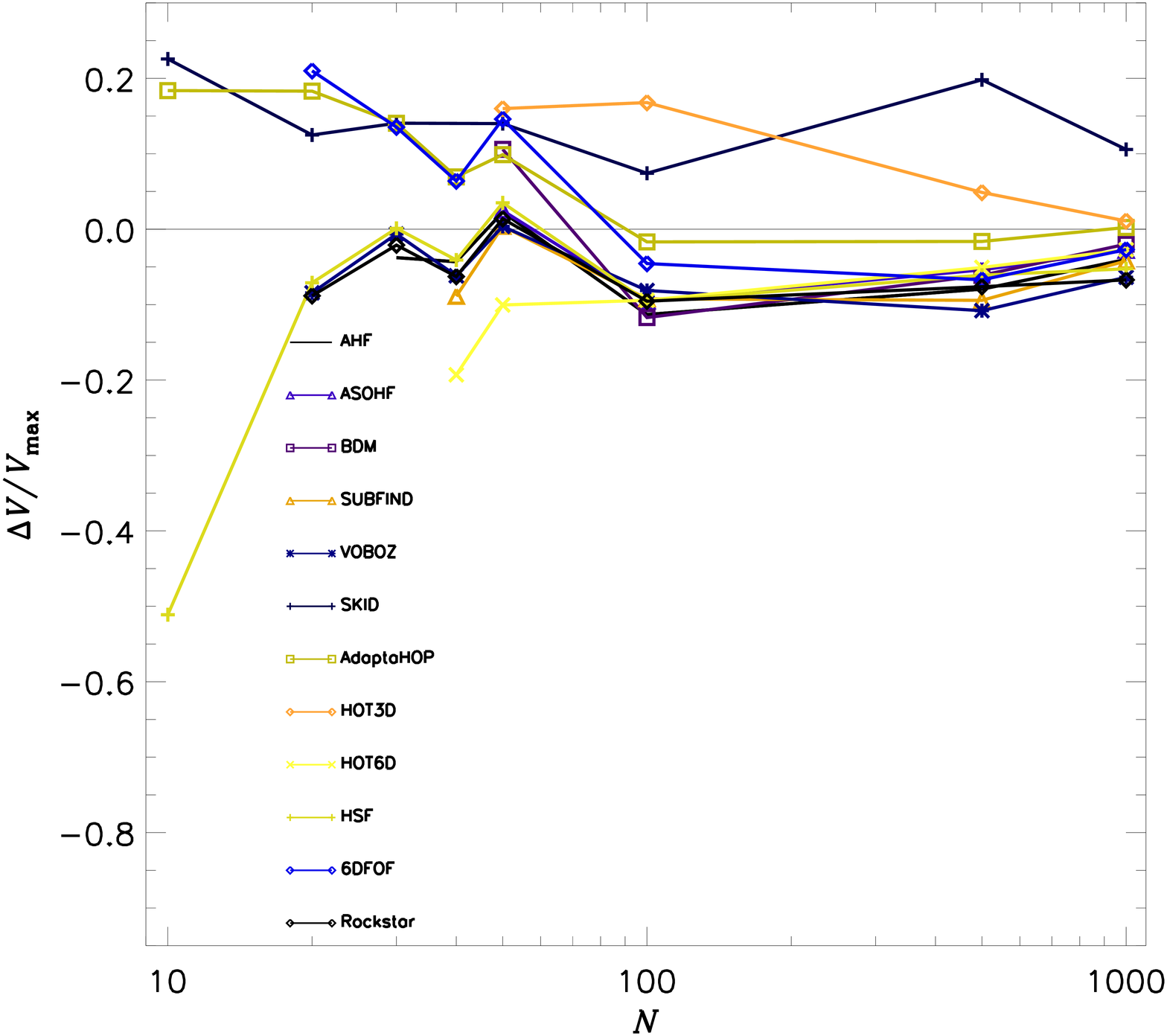,width=\hsize,angle=0}
  \caption{Fractional difference between theoretical maximum of the
    rotation curve and the numerically derived maximum vs. the
    theoretical maximum for the subhalo.}
\label{fig:RES-Vmax}
\end{figure}

\paragraph*{Maximum of Rotation Curve}
As the number of particles is merely a measure for the
cross-performance of halo finders and not (directly) related to
credible subhalo properties we also need to have a look at \vmax\
again. The fractional error as a function of the (theoretical) number
of subhalo particles is plotted in \Fig{fig:RES-Vmax}. We note that
aside from those halo finders who showed a contamination by host
particles all codes recover the theoretical maximum of the rotation
curve down to the limit of their subhalo's visibility (although
possibly the last data point for the lowest number of particles should
be discarded in that regard).

\subsubsection{The ``Blind Test''} \label{sec:blindtest}
Aside from the mock haloes analysed before we also designed a
particular test where none of the participants had foreknowledge of what it
contained; only Stuart Muldrew, who generated all the mock haloes,
knew the setup that is summarised in \Tab{tab:blindtest} where
the type ``host'' refers to the host halo and ``sub'' to a subhalo. We
dubbed this individual test the ``blind test''. Please note that some
of the subhalo's density profiles in this test followed a Hernquist model
\citep[][marked ``Hern'' in the Table,]{Hernquist90} instead of the NFW
profile. Further, two haloes were deliberately placed at the same
location yet with diametrically opposed velocities.

\begin{figure}
  \psfig{figure=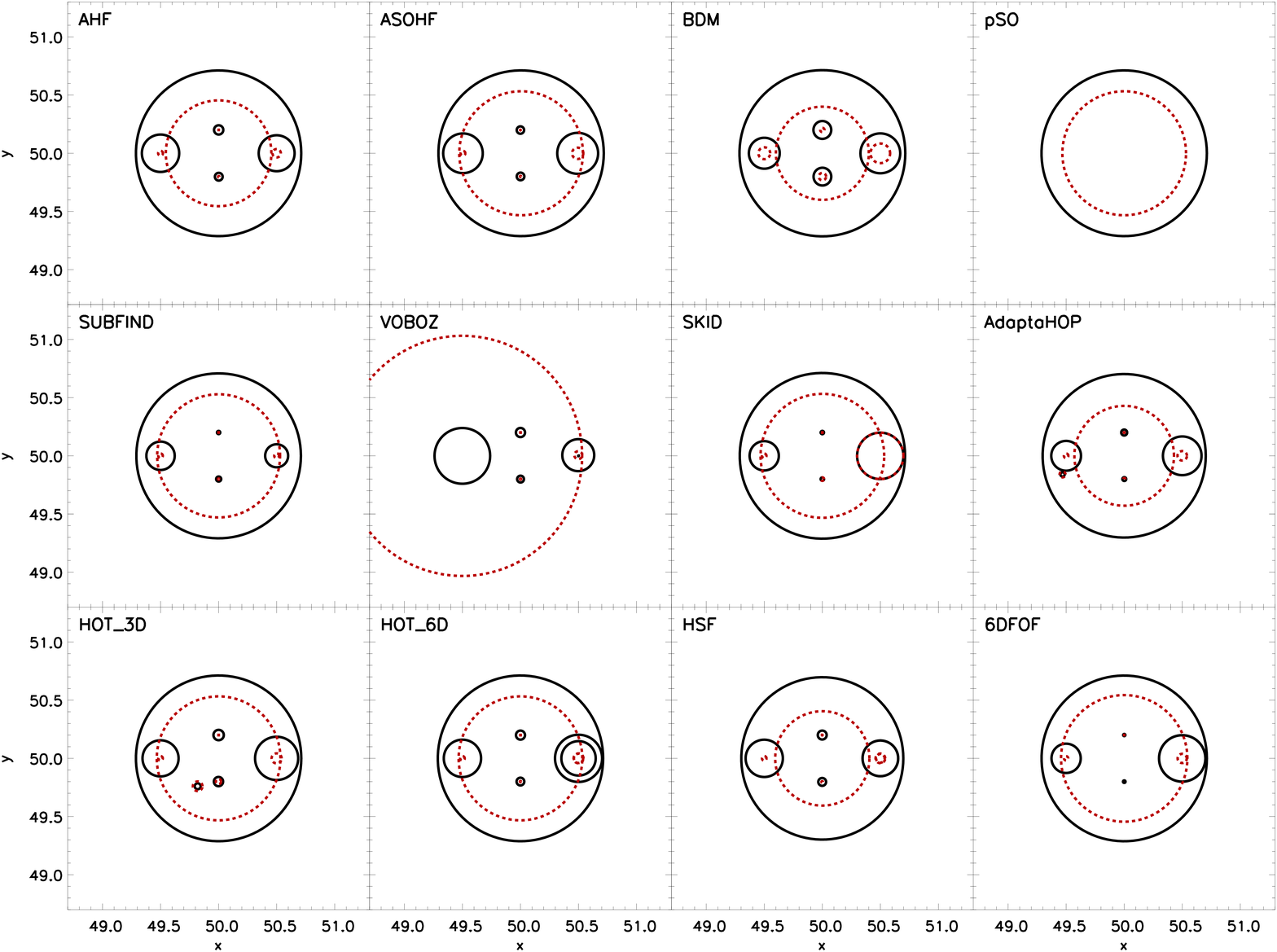,width=\hsize,angle=0}
  \caption{Visual impression of the ``blind test'' (projection into
    $x-y$ plane). Each halo found is represented by a circle with a
    radius equal to the fiducial $R_{200}$ value (solid black) and the
    $R_{\rm vmax}$ value (dashed red).}
\label{fig:blind-xy}
\end{figure}

As this test more or less marked the end of the workshop and was
primarily considered a fun exercise, we did not include it in the
actual data set presented in \Sec{sec:mockdata}. Please note that not
all halo finders participated and that we did not give the players in
the game a chance to tune their code parameters to the data
set. Nevertheless we decided to simply show visual impressions of
those who returned results in \Fig{fig:blind-xy}. There we merely show
the projections of the (fiducial) $R_{200}$ and $R_{\rm vmax}$ radii
in the $x-y$ plane as the $z$ coordinate of all haloes is identical.

\begin{table}
  \caption{Summary of the haloes in the blind test. Positions are
    given in \hMpc\ and velocities in km/sec.}
\label{tab:blindtest}
\begin{center}
\begin{tabular}{lllllcccl}
\hline
type & $N_{\rm 100}$ & $x$ & $y$ & $z$ & $v_x$ & $v_y$ & $v_z$ & profile\\
\hline
host & 10$^6$ & 50 & 50 & 50 & 0 & 0 & 0 & NFW\\
sub  & 10$^4$ & 50.5 & 50 & 50 & -10$^3$ & 0 & 0 & NFW\\
sub  & 10$^4$ & 50.5 & 50 & 50 & 10$^3$ & 0 & 0 & NFW\\
sub  & 10$^4$ & 49.5 & 50 & 50 & 10$^3$ & 0 & 0 & Hern\\
sub  & 10$^2$ & 50 & 49.8 & 50 & 10$^3$ & 10$^3$ & 0 & NFW\\
sub  & 10$^2$ & 50 & 50.2 & 50 & 0 & -10$^3$ & 0 & Hern\\
\hline
\end{tabular}
\end{center}
\end{table}

It is interesting to note that the phase-space halo finders were again
capable of locating the two overlapping subhaloes even though this is
not clearly visible in the projection (as their circles are obviously
overlapping). Of the 3D finders \texttt{SKID} noticed that there was
something odd at that position, returning one object with double the
mass (and $R_{\rm vmax}$ extending out to the outer radius). All other
halo finders only found one of the two subhaloes. Also remember that
\texttt{pSO} is not (yet) designed to find subhaloes and hence only
the host has been returned. It is further remarkable that none of the
halo finders had trouble finding the two small subhaloes while the
host had not been found for some of the codes.

Again, we would like to stress that this test should not be taken too
seriously. However, we nevertheless remark that analysing a
cosmological simulation is also a sort of ``blind'' analysis as the
answer is not previously known.

\subsection{Cosmological Simulation} \label{sec:cosmologicalsimulation}

\begin{figure*}
  \psfig{figure=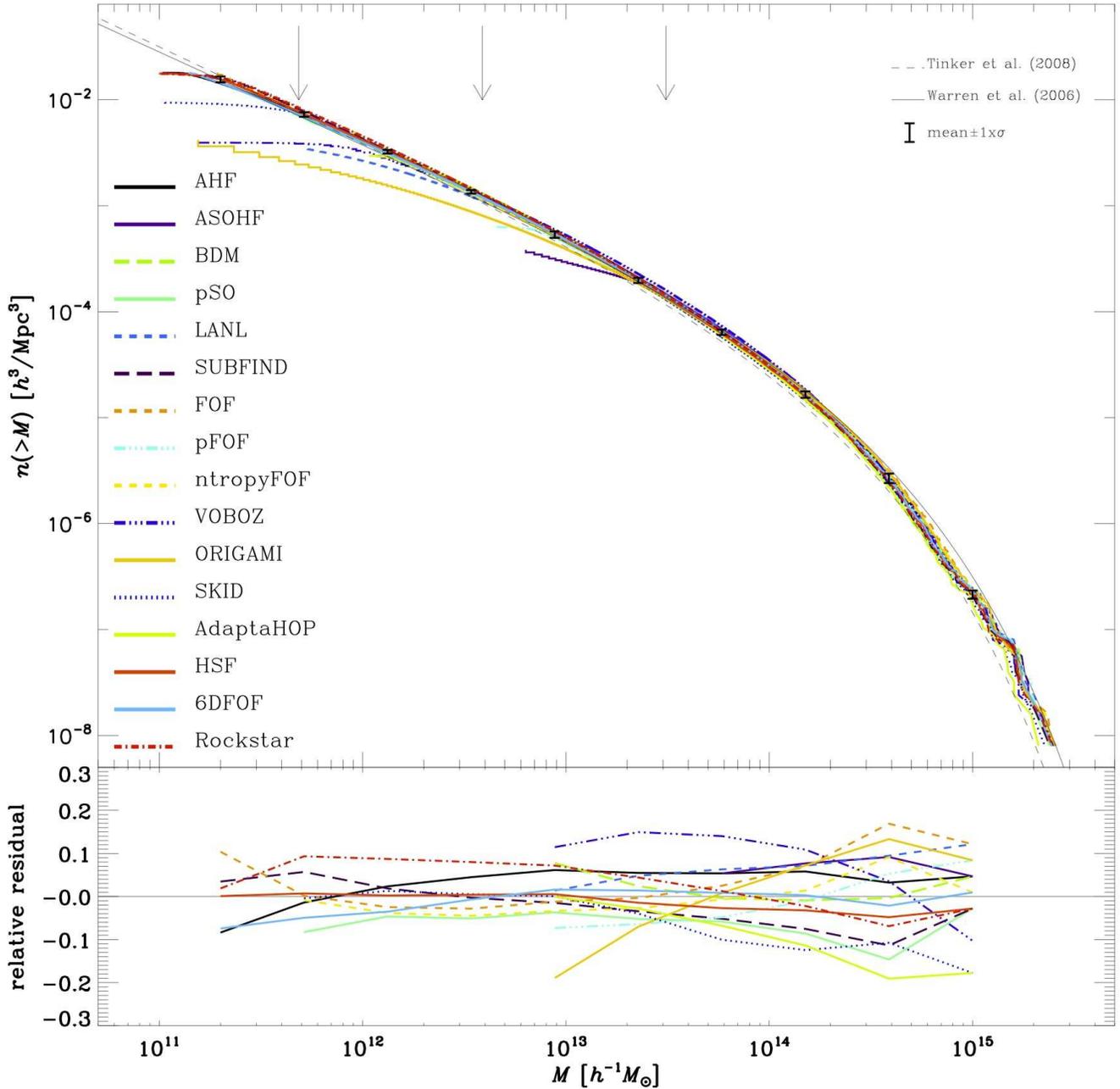,width=\hsize,angle=0}
  \caption{Upper panel: the cumulative mass ($M_{200}$) function. The
    arrows indicate the 50 particle limit for the 1024$^3$ (left),
    512$^3$ (middle), and 256$^3$ (right) simulation data. The thin
    black lines crossing the whole plot corresponds to the mass
    function as determined by \citet[][(solid)]{Warren06} and
    \citet[][(dashed)]{Tinker08}. The error bars represent the mean
    mass function of the codes ($\pm 1 \sigma$).  Lower panel: the
    fractional difference of the mean and code halo mass functions.
    For more details please refer to the text.}
\label{fig:massfunc}
\end{figure*}

We now turn to the comparison of a real cosmological simulation
including a substantial number of objects formed and embedded within
the large-scale structure of the Universe.

However, even though the simulation contains a large number of
particles (i.e. up to 1024$^3$ in the highest resolved data set) the
given volume of side length 500\hMpc\ does not allow for a study of
subhaloes in detail: for the fiducial 512$^3$ particle run the largest
object in the simulation box merely contains of order 10 subhaloes
with the number of substructure objects dramatically decreasing when
moving to (potentially) lower mass host haloes. We therefore stress that
this particular comparison only focuses on field haloes and hence is
well suited even for those codes that (presently) cannot cope with
subhaloes.

Further, as mentioned already in \Sec{sec:cosmodata} we have the data
available at various resolutions ranging from 256$^3$ to 1024$^3$
particles. We decided to use the highest resolution analysis performed
by each finder as has already been summarised in \Tab{tab:codes} in the
subsequent comparison plots. The analysis in this particular Section
primarily revolves around the (statistical) recovery of halo
properties. In that regard we are nevertheless limiting our analysis
to properties akin to the ones already studied in
\Sec{sec:mockhaloes}, namely the mass $M$, the position $\vec{R}$, the
peak of the rotation curve \vmax, and the (bulk) velocity $V_{\rm
bulk}$. We are going to utilise masses as defined via
$200\times\rho_{\rm crit}$, i.e. $M_{\rm 200}$.

We like to re-iterate at this point again that for this particular
comparison each halo finder returned halo properties as derived from
applying the code to the actual data set; we aim at comparing the
results of the codes for each and every single one being applied to
the data individually. We consider this the most realistic comparison
as this directly gauges the differences of the resulting halo
catalogues.

We have already seen that all halo finders are capable of recovering
the mass of mock haloes, irrespective of whether the density profile
is cored or has a cusp (cf. \Fig{fig:MOCK-M200}). We therefore do not
expect to find surprising differences in the first and most obvious
comparison, i.e. the (cumulative) mass function presented in
\Fig{fig:massfunc}. Please note that \texttt{pFOF}
discarded objects below 100 particles and hence did not return haloes
below $\approx 8\times 10^{12}$\hMsun; similarly, \texttt{pSO}
discarded objects with fewer than 50 particles, according to the
criterion laid out in equation~(30) of \citet{Lukic07}. And in each
case the (cumulative) mass function starts to flatten at approximately
the resolution limit of the simulation analysed by the respective
code.

However, \texttt{ORIGAMI} seems to miss some low-mass structures
caught by other halo-finders.  One possible reason is that some
smaller density enhancements seen by other finders have not undergone
shell-crossing along three axes, and therefore do not meet
\texttt{ORIGAMI}'s definition of a halo.  Another is that
\texttt{ORIGAMI} may be missing many subhaloes, which it does not
attempt to separate from parent haloes.

Further, the \texttt{LANL} halo finder is designed to be an FOF finder
and, if needed, SO objects are defined on top of such
friends-of-friends haloes.  Thus, for smaller haloes completeness is
an issue as not every SO halo will have an FOF counterpart. Of course,
it is possible to run the code in the limit $b \rightarrow 0$ and
$N_{min}=1$, having each particle serving as a potential centre of an
SO halo, but the increase in computational cost would make this
impractical, as direct SO halo finders which do precisely this in a
more effective manner already exist. Nevertheless, we can see that
computationally very fast method of growing SO spheres on top of FOF
proxy haloes result in excellent match when compared to direct SO
finders for well sampled haloes ($\sim$500 particles per halo).

In order to better view (possible) differences in the mass functions
we further calculated the ``mean mass function'' in 10 logarithmically
placed bins across the range $2\times 10^{11}$~--~$1\times
10^{15}$\hMsun\ alongside $1\sigma$ error bars for the means. Note
that all codes only contributed to those bins where their data set is
considered complete. We further deliberately stopped the binning at
$1\times 10^{15}$\hMsun\ to not be dominated by small number
statistics for the few largest objects. The results can also be viewed
in \Fig{fig:massfunc} too, where we also show in the bottom panel the
fractional difference between the mean and the code mass functions
across the respective mass range. And we additionally added as thin
solid black line to the actual mass function plot in the upper panel
of \Fig{fig:massfunc} the numerically determined mass function of
\citet{Warren06} which is based upon a suite of sixteen 1024$^3$
simulations of the $\Lambda$CDM universe as well as the one derived by
\citet{Tinker08} derived from a substantial set of cosmological
simulations actually including the ones used by \citet{Warren06}
(cf. their Fig.1). Note that the former is based upon FOF and the
latter on SO masses.

\begin{figure}
  \psfig{figure=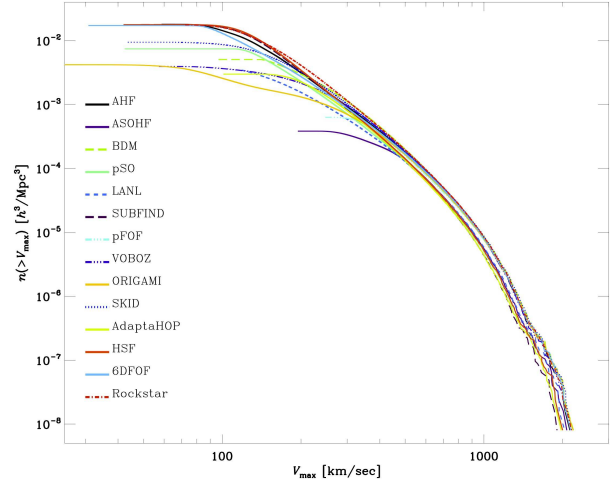,width=\hsize,angle=0}
  \caption{The cumulative \vmax\ function.}
\label{fig:vmaxfunc}
\end{figure}

As highlighted in the Introduction~\ref{sec:howtocomparehaloes} the
peak value of the rotation curve may be a more suitable quantity to
use when it comes to comparing the masses of (dark matter) haloes. We
therefore show in the \Fig{fig:vmaxfunc} the cumulative distribution
of \vmax. Apart from the expected flattening at low \vmax\ due to
resolution we now note that this is in fact the case: codes that did
not estimate masses according to the standard definition $M(<R)=
4\pi/3 \ R^3 \ \Delta\rho $ nevertheless recovered the correct \vmax\
values. Given the ability of comparing \vmax\ to observational data
(cf. \Sec{sec:howtocomparehaloes}) we conclude that \vmax\ is a more
meaningful quantity which can serve as a proxy for mass. Please note
again the flattening of some curves at the low-\vmax\ end due to
either the resolution of the simulation analysed or an imposed minimum
number of particles cut and that not all FOF-based finders
  returned a \vmax\ value.

\begin{figure}
  \psfig{figure=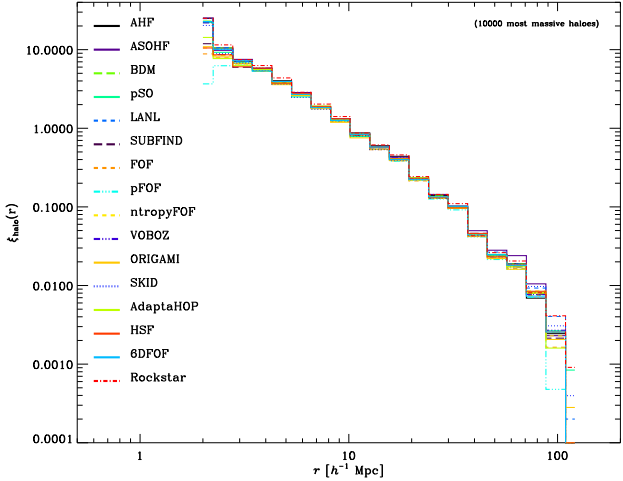,width=\hsize,angle=0}
  \caption{The 2-point correlation function for the 10000 most massive
    objects.}
\label{fig:xi}
\end{figure}

We have seen in \Sec{sec:mockhaloes} that there exists some scatter
between halo finders in the recovery of the halo position. It
therefore appears mandatory to check for differences in halo positions
recovered from the cosmological simulation, too. To this extent we
calculated the 2-point correlation function and present the results in
\Fig{fig:xi}. In order to analyse a comparable data set (remember that
some codes analysed the 1024$^3$, some the 512$^3$, and some the
256$^3$ particle simulation) we restricted the haloes to the 10000
most massive objects and found excellent agreement.\footnote{Please
  note that it makes little difference to use the 10000 objects with
  the largest \vmax\ value as there is a strong correlation between
  $M$ and \vmax\ for each code. In the end we are interested in
  limiting the analyses to the $N$ most massive objects and hence a
  ``mis-calculation'' of the mass is irrelevant as long as differences
  in mass are systematic as in our case.} The smallest scale
considered in this comparison is 2\hMpc\ in order not to probe the
interiors of galaxy clusters. The minute drop of
the correlation function for \texttt{pFOF} at the smallest scale
probed may be explained by the usage of the marginally larger linking
length of $b=0.2$ applied during their analysis and the fact that
\texttt{pFOF} uses the centre of mass instead of the density peak as
the centre of the halo.

\begin{figure}
  \psfig{figure=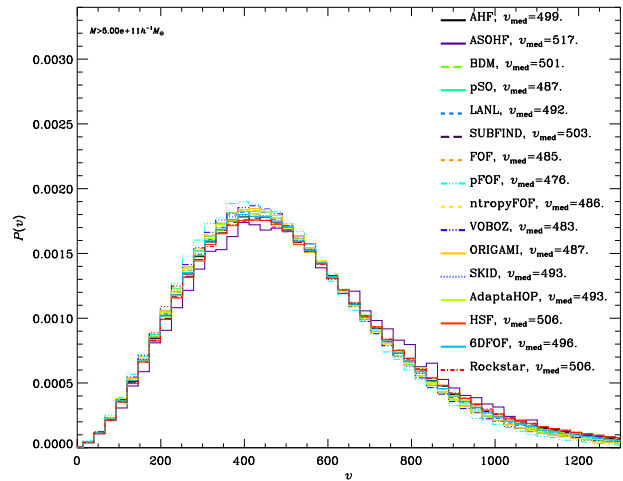,width=\hsize,angle=0}
  \caption{The distribution of bulk velocities for objects more
    massive than $5\times 10^{11}$\hMsun.}
\label{fig:Pvel}
\end{figure}

Finally we cross-compare the bulk velocities of haloes in
\Fig{fig:Pvel} where we find excellent agreement. We further give in
the legend the medians of the distribution for each halo finder: the
mean (of the medians) is 489 km/sec with a $1-\sigma$ of 9 km/sec
(i.e. 2\% deviation).

\section{Summary \& Conclusions} \label{sec:conclusions}
We have performed an exhaustive comparison of 18 halo finders for
cosmological simulations. These codes were subjected to various suites
of test scenarios all aimed at addressing issues
related to the subject of identifying gravitationally bound objects in
such simulations.

The tests consisted of idealized mock haloes set up according to a
specific matter density profile (i.e. NFW and Plummer) where we
studied isolated haloes as well as (sub-)subhaloes. We further
utilized a cosmological simulation of the large-scale structure of the
universe primarily containing field haloes. The requirement for the
mock haloes was to simply return the centres of the identified objects
alongside a list of particles (possibly) belonging to that halo. We
then applied a universal tool to calculate all other quantities
(e.g. bulk velocity, rotation curve, (virial) mass, etc.). For the
cosmological data the code representatives were simply asked to return
their ``best'' values for a suite of canonical values.

\paragraph*{Mock Haloes}
We found that the deviation of the recovered position to the actual
centre of the object is largest for FOF-based methods which is
naturally explained by the fact that they define centres as
centre-of-mass whereas most other codes identify a peak in the density
field. Further, dark matter haloes that have an intrinsic core (e.g. a
Plummer sphere) yield larger differences between the input centre and
the recovered centre for most codes. Such density profiles are not
expected within the Universe we inhabit. However, the bulk velocities,
(virial) masses, and \vmax\ values satisfactorily agreed with the
analytical input irrespective of the underlying density profile -- at
least for host and subhaloes; sub-subhaloes still showed at times
departures as large as 50\% in mass and 20\% for \vmax. Please
note that all results are based upon the same post-processing software
and only the list of particles (and the centre) were determined by
each halo finder individually. Hence, variations in the centre will
automatically lead to differences as both mass and rotation curve are
spherically averaged quantities.

We further investigated the dependence of subhalo properties upon the
position within the host, in particular its distance to the
centre. There we found that -- while all codes participating in this
exercise recovered excellent \vmax\ values for a NFW subhalo sampled
with 10000 particles inside a NFW host two orders of magnitude more
massive\footnote{Note that only halo finders capable of identifying
substructure can participate in a comparison of (sub-)subhalo
properties.} -- phase-space finders excelled by also locating the
subhalo when it overlapped with the centre of the host. However, in
this case they struggle to properly calculate its properties.

Putting a subhalo at varying positions inside a host is closely
related to a subhalo actually falling into a host. However, the latter
also introduces distortions in the shape of the subhalo due to tidal
forces while it is plunging through the background potential of the
host. We performed a simulation of the scenario where a subhalo
initially containing 10000 particles shoots right through the centre
of a host two orders of magnitude more massive. While we found that
the number of particles significantly drops when the subhalo
approaches the host's centre, it rises again to a plateau level after
the central passage -- and this is apparent in all codes. The peak of
the rotation curve, which should be less susceptible to (tidally
induced) variations in the outer subhalo regions, shows less
variation. However, \vmax\ actually rises shortly after the subhalo
leaves the very central region indicative of two (related) effects:
contamination with host particles and problems with the unbinding
procedure. Nevertheless, these problems are (still) common to all halo
finders used in this particular study and they all mutually agree upon
the initial and final value.

Another question addressed during our tests with the mock haloes was
the number of particles required in a subhalo in order to still be
able to separate it from the host background.  To this extent we
successively lowered the number of particles used to sample a subhalo
that had been placed at half the $M_{100}$ radius of the host. We
found that the majority of finders participating in this exercise are
capable of identifying the subhalo down to 30-40 particles. Yet again,
(most of) the phase-space finders even locate the object with as few
as 10-20 particles. Some of the configuration space finders also
tracked down the subhalo to such low numbers of particles, however,
they did not obtain the correct particle lists leading to subhalo
properties that differ from the analytical input values.

We would like to close this part of the summary with the notion that
while there is a straight-forward relation between (virial) mass and
the peak of the rotation curve for isolated field haloes (once the
density profile is known), the mass of a subhalo is more ambiguously
defined. As we have seen, it is (in most situations) more meaningful
to utilize the peak of the rotation curve as a proxy for mass
(cf. \Fig{fig:MOCK-M200Dist} vs. \Fig{fig:MOCK-VmaxDist} as well as
\Fig{fig:DYN-Npart} vs. \Fig{fig:DYN-Vmax}). However, as could
  also be witnessed in \Fig{fig:DYN-Vmax}, quite a number of halo
  finding techniques gave rise to an artificial increase of \vmax\
  right after the passage through the centre of its host obscuring its
  applicability as a mass representative.

\paragraph*{Cosmological Simulation}
As a matter of fact there is little to say regarding the comparison of
the cosmological data set; as can be seen in Figs.~\ref{fig:massfunc}
through~\ref{fig:Pvel} the agreement is well within the (omitted)
error bars for the basic properties investigated here (i.e. mass,
velocity, position, and \vmax). And unless we can be certain
  which halo finding technique is the ultimate (if such exists at
  all), the observed scatter indicates the accuracy to which we can
  determine these properties in cosmological simulations. We would
though like to caution that the haloes found within the cosmological
simulation are primarily well defined and isolated objects and hence
it is no surprise that we find such an agreement. Subhaloes, however,
are not well defined and therefore lead to larger differences between
halo finders as seen during the comparison of the mock haloes. For
those codes that diverge from the general agreement the differences
are readily explained and have been discussed in
\Sec{sec:cosmologicalsimulation}.

\paragraph*{Concluding Remarks}

The agreement amongst the different codes is rather remarkable and
reassuring. While they are based upon different techniques and -- even
for those based upon same techniques -- different technical parameters
they appear to recover comparable properties for dark matter haloes as
found in state-of-the-art simulations of cosmic structure
formation. We nevertheless need to acknowledge that some codes require
improvement.  For instance, phase-space finders find halo centres even
if the centre overlaps with another (distinct) object and recover
subhaloes to smaller particle number, however they still have problems
with the (separated) issue of assigning the correct particles in these
cases and hence deriving halo properties afterwards.

We close with the remark that we deliberately did not dwell on the
actual technical parameters of each and every halo finder as this is
beyond the scope of this paper and we refer the reader to the
respective code papers for this. However, it is important to note that
with an appropriate choice of these parameters the results can be
brought into agreement. This is an important message from this
particular study. We are not claiming that all halo finders need to
return identical results, but they can (possibly) be tuned that
way. In that regards we also like to remind the reader again that this
particular comparison is aimed at comparing codes as opposed to
algorithms: we even tried to gauge the differences found when applying
codes based upon the same algorithm to identical data sets.

\section*{Acknowledgements}

We are greatly indebted to the ASTROSIM network of the European
Science Foundation (Science Meeting 2910) for financially supporting
the workshop ``Haloes going MAD'' held in Miraflores de la Sierra near
Madrid in May 2010 where all of this work was initiated.

AK is supported by the {\it Spanish Ministerio de Ciencia e
  Innovaci\'on} (MICINN) in Spain through the Ramon y Cajal programme
as well as the grants AYA 2009-13875-C03-02, AYA2009-12792-C03-03,
CSD2009-00064, and CAM S2009/ESP-1496. He further thanks Lee Hazlewood
for summer wine.  SRK acknowledges support by the MICINN under the
Consolider-Ingenio, SyeC project CSD-2007-00050.  SP and VQ have also
been supported by the MICINN (grants AYA2010-21322-C03-01 and
CONSOLIDER2007-00050) and the {\it Generalitat Valenciana} (grant
PROMETEO-2009-103). SP also thanks the MICINN for a FPU doctoral
fellowship.  MZ is supported by NSF grant AST-0708087. JD is supported
by the Swiss National Science Foundation.  MAA, BLF, and MCN are
grateful for discussions with and support from Alex Szalay, and
funding from the W.M. Keck and Gordon and Betty Moore Foundations.
PMS acknowledges support under a DOE Computational Science Graduate
Fellowship (DE-FG02- 97ER25308). The software used by PMS and PMR in
this work was in part developed by the DOE-supported ASC/Alliance
Center for Astrophysical Thermonuclear Flashes at the University of
Chicago. Further, PMS and PMR used resources of the National Center
for Computational Sciences at Oak Ridge National Laboratory, which is
supported by the Office of Science of the US Department of Energy
under contract no. DE-AC05-00OR22725.  JIR would like to acknowledge
support from SNF grant PP00P2\_128540 / 1.  SG and VT acknowledge
support by the Deutsche Forschungsgemeinschaft (DFG).  KD acknowledges
the support by the DFG Priority Programme 1177 and additional support
by the DFG Cluster of Excellence ``Origin and Structure of the
Universe''.  The work was done while CH was working for Los Alamos
National Laboratory.  Part of the work was supported by the DOE under
contract W-7405-ENG-36.  CH, PF, and ZL acknowledge support from the
LDRD program at Los Alamos National Laboratory. ZL was supported in
part by NASA. A special acknowledgment is due to supercomputing time
awarded to us under the LANL Institutional Computing Initiative.  GY
acknowledges financial support from MICINN (Spain) under project AYA
2009-13875-C03-02 and the ASTROMADRID project S2009/ESP-1496 financed
by Comunidad de Madrid. PSB received support from the U.S. Department
of Energy under contract number DE-AC02-76SF00515.

\bibliographystyle{mn2e}
\bibliography{archive}

\bsp

\label{lastpage}

\end{document}